\documentclass[twocolumn,aps,pra,showpacs,tightenlines]{revtex4-1}
\usepackage{amsmath}
\usepackage{amsfonts}
\usepackage{graphicx}
\usepackage{epsfig}
\usepackage{color}
\usepackage[colorlinks,citecolor=blue]{hyperref}
\usepackage{multirow}
\newcommand{\tabincell}[2]{\begin{tabular}{@{}#1@{}}#2\end{tabular}}

\begin{document}
\title{Enhancement of few-photon optomechanical effects with cross-Kerr nonlinearity}

\author{Fen Zou}
\affiliation{Key Laboratory of Low-Dimensional Quantum Structures and
Quantum Control of Ministry of Education, Department of Physics and
Synergetic Innovation Center for Quantum Effects and Applications, Hunan
Normal University, Changsha 410081, China}

\author{Li-Bao Fan}
\affiliation{Key Laboratory of Low-Dimensional Quantum Structures and
Quantum Control of Ministry of Education, Department of Physics and
Synergetic Innovation Center for Quantum Effects and Applications, Hunan
Normal University, Changsha 410081, China}

\author{Jin-Feng Huang}
\email{jfhuang@hunnu.edu.cn}
\affiliation{Key Laboratory of Low-Dimensional Quantum Structures and
Quantum Control of Ministry of Education, Department of Physics and
Synergetic Innovation Center for Quantum Effects and Applications, Hunan
Normal University, Changsha 410081, China}

\author{Jie-Qiao Liao}
\email{jqliao@hunnu.edu.cn}
\affiliation{Key Laboratory of Low-Dimensional Quantum Structures and
Quantum Control of Ministry of Education, Department of Physics and
Synergetic Innovation Center for Quantum Effects and Applications, Hunan
Normal University, Changsha 410081, China}

\date{\today}

\begin{abstract}
Few-photon optomechanical effects are not only important physical evidences for understanding the radiation-pressure interaction between photons and mechanical oscillation, but also have wide potential applications in modern quantum technology. Here we study the few-photon optomechanical effects including photon blockade and generation of the Schr\"{o}dinger cat states under the assistance of a cross-Kerr interaction, which is an inherent interaction accompanied the optomechanical coupling in a generalized optomechanical system. By exactly diagonalizing the generalized optomechanical Hamiltonian and calculating its unitary evolution operator, we find the physical mechanism of the enhancement of photon blockade and single-photon mechanical displacement. The quantum properties in this generalized optomechanical system are studied by investigating the second-order correlation function of the cavity field and calculating the Wigner function and the probability distribution of the rotated quadrature operator for the mechanical mode. We also study the influence of the dissipations on the few-photon optomechanical effects.
\end{abstract}
\maketitle


\section{Introduction}

The radiation-pressure interaction between the optical and mechanical degrees of freedom is at the center of the field of cavity optomechanics~\cite{Vahala2008Science,Meystre2012PT,Aspelmeyer2014RMP}. This interaction takes the form as a trilinear two-mode coupling, which describes that the cavity photons exert a photon-number dependent force on a mechanical oscillator~\cite{Law1995PRA}.
According to the magnitude of the optomechanical coupling, people usually consider two kinds of regime of the optomechanical coupling: (i) the many-photon involved coupling case and (ii) the few-photon involved coupling case. Generally, the physical phenomena involved many photons is easily observed because the coupling between the photons and phonons is effectively enhanced by a factor of the square root of the cavity photon number under the linearization frame. In the many-photon coupling case, many advances have been made in relating topics such as normal-mode splitting~\cite{Dobrindt2008PRL,Groblacher2009Nature,Teufel2011NatureA,Verhagen2012Nature}, quantum cooling of mechanical resonators~\cite{Wilson-Rae2007PRL,Marquardt2007PRL,Genes2008PRA,Teufel2011Nature,Chan2011Nature, Li2008PRB,Liu2013PRL,Lai2018PRA}, optomechanical entanglement~\cite{Vitali2007PRL,Hartmann2008PRL,Tian2013PRL,Wang2013PRL,Palomaki2013Science}, optomechanically induced transparency~\cite{Agarwal2010PRA,Weis2010Science,Safavinaeini2011Nature}, and generation of squeezed light~\cite{Brooks2012Nature,Naeini2013Nature,Purdy2013PRX}.

In contrast, to observe the evidence of the optomechanical coupling at the few-photon level, the single-photon optomechanical-coupling strength is required to be sufficiently large so that the physical phenomenon induced by a single photon can be resolved from the noise in this system~\cite{Rabl2011PRL,Nunnenkamp2011PRL,Liao2012PRA,Hong2013PRA,Liao2013PRA,Liu2013PRA}. The observation of photon blockade effect~\cite{Imamoglu1997PRL,Birnbaum2005nature,Liew2010PRL,Rabl2011PRL,Bamba2011PRA,Nori2013PRA,Huang2013PRA,Huang2018PRL} is an important task in cavity optomechanics working in the single-photon strong-coupling regime, where the single-photon optomechanical-coupling strength is larger than the cavity-field decay rate. Another interesting task in the few-photon optomechanics is the generation of the mechanical Schr\"{o}dinger cat states~\cite{Marshall2003PRL,Tian2005PRB,Isart2011PRL,Pepper2012PRL,Yin2013PRA,Tan2013PRA,Ge2015PRA,Liao2016PRA,Liao2016PRL}, which is based on the conditional displacement mechanism of the optomechanical coupling. To create quantum superposition of distinct coherent states, the single-photon optomechanical-coupling strength should be larger than the resonance frequency of the mechanical mode~\cite{Marshall2003PRL}. Recently, some proposals have been proposed to enhance the mechanical displacement induced by single photons~\cite{Liao2016PRL,Liao2015PRA}.

The above mentioned two tasks require a sufficiently large optomechanical coupling. However, the single-photon strong-coupling regime has not been realized in current experiments. Nevertheless, many proposals have been proposed to enhance the optomechanical coupling~\cite{Xuereb2012PRL,Rimberg2014NJP,Heikkila2014PRL,Pirkkalainen2015NC,Liao2014NJP,Lue2015PRL,Lemonde2016NC} such that the systems reach the single-photon strong-coupling regime. For example, in Ref.~\cite{Heikkila2014PRL} the authors proposed to enhance the single-photon optomechanical coupling by utilizing the nonlinearity of the Josephson junctions. With this method, the single-photon optomechanical coupling can be enhanced several orders of magnitude. Meanwhile, this circuit also creates a cross-Kerr interaction~\cite{Heikkila2014PRL,Heikkila2015PRA,You2016PRA,Sarma2017JOS} between the optical mode and the mechanical mode, and the magnitude of the cross-Kerr interaction might be a fraction of the single-photon optomechanical coupling strength. Based on the fact that the \emph{original motivation} of the proposal based on the superconducting circuit is to \emph{enhance the single-photon optomechanical coupling} and to further realize the few-photon optomechanical tasks, it is therefore natural to ask the question: what is the effect of the additional cross-Kerr interaction on the \emph{few-photon optomechanical tasks} such as the photon blockade and the generation of the Schr\"{o}dinger cat states?

In this paper, we study the photon blockade effect and the generation of quantum superposition of coherent states in a superconducting quantum circuit proposed in Ref.~\cite{Heikkila2014PRL}. In particular, we will focus on the effect of the cross-Kerr interaction on the photon blockade and the mechanical cat state generation. We will analyze the effective photon nonlinear interaction induced by the optomechanical coupling and the cross-Kerr interaction by calculating the equal-time second-order correlation function of the cavity photons. In this case, the cross frequency modulation will change the effective photon nonlinearity and the photon blockade. We will also analyze the effect of the cross-Kerr interaction on the magnitude of the single-photon mechanical displacement. The cross frequency modulation will change effectively the driving detuning of the single photon and hence
change the magnitude of the mechanical displacement.

The rest of this paper is organized as follows. In Sec.~\ref{modelsec}, we introduce the physical model and present the Hamiltonian. In Sec.~\ref{photonblockade}, we study the effect of the cross-Kerr interaction on the photon blockade effect in the optomechanical cavity. In Sec.~\ref{Schcatstate}, we study the generation of Schr\"{o}dinger-cat states in the mechanical mode and investigate the Wigner function and the probability distribution of the rotated quadrature operator to analyze the quantum interference and coherence effects in the generated cat states. We also study the influence of the optical and mechanical dissipations on the cat-state generation. Finally, we present some discussions on the experimental implementation of this model and conclude this work in Sec.~\ref{conclusion}. A detailed derivation of the unitary evolution operator associated with the generalized optomechanical Hamiltonian is presented in the Appendix.

\section{Model and Hamiltonian \label{modelsec}}

\begin{figure}[tbp]
\center
\includegraphics[bb=0 0 541 505, width=0.47 \textwidth]{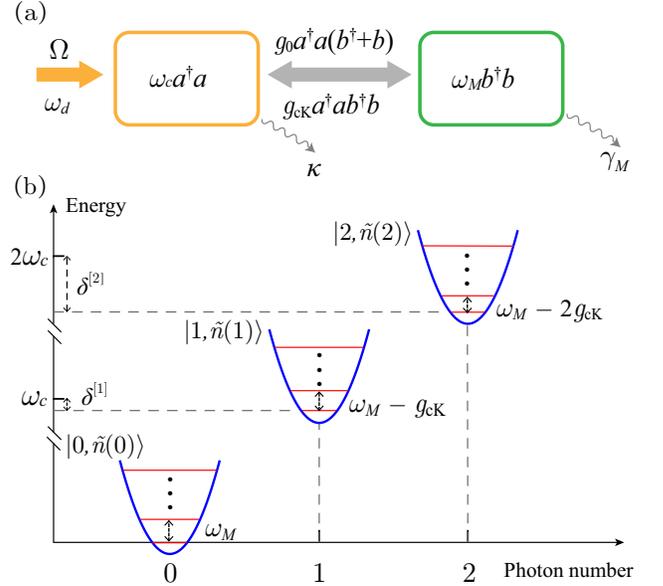}
\caption{(Color online) (a) Schematic diagram of the generalized optomechanical model, which is composed of a single cavity-field mode and a single mechanical mode. The two modes are coupled to each other through both optomechanical and cross-Kerr interactions. (b) Diagram of the eigenenergy spectrum of the Hamiltonian $\hat{H}_{\text{gom}}$ in the subspace associated with zero, one, and two photons.}
\label{Fig1}
\end{figure}
We consider a generalized optomechanical model, which is composed of a single-mode optical field and a mechanical mode [see Fig.~\ref{Fig1}(a)]. Here the optical mode is coupled to the mechanical mode via both the optomechanical interaction and the cross-Kerr interaction. The Hamiltonian of the generalized optomechanical model reads ($\hbar=1$)
\begin{equation}
\hat{H}_{\text{gom}}=\omega_{c}\hat{a}^{\dagger}\hat{a}+\omega_{M}\hat{b}^{\dagger}\hat{b}-g_{0}\hat{a}^{\dagger}\hat{a}(\hat{b}^{\dagger}+\hat{b})-g_{\text{cK}}\hat{a}^{\dagger}\hat{a}\hat{b}^{\dagger}\hat{b},\label{hams}
\end{equation}
where $\hat{a}$ ($\hat{a}^{\dagger}$) and $\hat{b}$ ($\hat{b}^{\dagger}$) are, respectively, the annihilation (creation) operators of the cavity mode and the mechanical mode, with the corresponding resonance frequencies $\omega_{c}$ and $\omega_{M}$. The $g_{0}$ term denotes the optomechanical coupling between the cavity field and the mechanical mode~\cite{Law1995PRA}, with $g_{0}$ being the single-photon optomechanical-coupling strength. The $g_{\text{cK}}$ term describes the cross-Kerr interaction between the cavity field and the mechanical mode~\cite{Heikkila2014PRL}, with the coupling strength $g_{\text{cK}}$. Note that this model has been realized in an electromechanical system which is proposed to enhance the single-photon optomechanical coupling by utilizing the nonlinearity of the Josephson junctions in Ref.~\cite{Heikkila2014PRL}. The cross-Kerr interaction is a by-product coupling in the hybrid system consisting of a superconducting qubit coupled to both a superconducting resonator and a mechanical resonator.

The photon number operator $\hat{a}^{\dagger}\hat{a}$ in the generalized optomechanical Hamiltonian $\hat{H}_{\text{gom}}$ is a conserved quantity due to $[\hat{a}^{\dagger}\hat{a},\hat{H}_{\text{gom}}]=0$. For a given photon number $m$, the Hamiltonian $\hat{H}_{\text{gom}}$ is reduced to a Hamiltonian describing a displaced harmonic oscillator of mode $\hat{b}$. In particular, the displacement force acting on the mechanical resonator is proportional to $mg_{0}$, and the resonance frequency of mode $\hat{b}$ is normalized to be $\omega_{M}-mg_{\text{cK}}$, with a photon-number-dependent frequency shift.

To calculate the eigensystem of $\hat{H}_{\text{gom}}$, we introduce a conditional displacement operator $\hat{D}(\hat{\xi})=\exp[\hat{\xi}(\hat{b}^{\dagger}-\hat{b})]$, where the displacement amplitude $\hat{\xi}$ is a nonlinear function of the photon number operator $\hat{a}^{\dagger}\hat{a}$,
\begin{equation}
\hat{\xi}=\frac{g_{0}\hat{a}^{\dagger}\hat{a}}{\omega_{M}-g_{\text{cK}}\hat{a}^{\dagger}\hat{a}}=\sum_{m=0}^{\infty}\xi^{[m]}\vert m\rangle_{a}\,_{a}\langle m\vert,\label{definbeta}
\end{equation}
with the $m$-photon induced mechanical displacement
\begin{equation}
\xi^{[m]}=\frac{mg_{0}}{\omega_{M}-mg_{\text{cK}}},
\end{equation}
where we introduce the number states $\vert m\rangle_{a}$ ($m = 0,1,2,\cdots$) of the cavity mode.
The Hamiltonian $\hat{H}_{\text{gom}}$ can be diagonalized as follows,
\begin{eqnarray}
\hat{\tilde{H}}_{\text{gom}} &=&\hat{D}^{\dagger}(\hat{\xi})\hat{H}_{\text{gom}}\hat{D}(\hat{\xi})\notag \\
&=&\omega_{c}\hat{a}^{\dagger}\hat{a}+(\omega_{M}-g_{\text{cK}}\hat{a}^{\dagger}\hat{a})\hat{b}^{\dagger}\hat{b}-\hat{\delta},
\end{eqnarray}
where we introduce the optical nonlinearity as
\begin{equation}
\hat{\delta}=\frac{g_{0}^{2}\hat{a}^{\dagger}\hat{a}\hat{a}^{\dagger }\hat{a}}{\omega_{M}-g_{\text{cK}}\hat{a}^{\dagger}\hat{a}}=\sum_{m=0}^{\infty}\delta^{[m]}\vert m\rangle_{a}\,_{a}\langle m\vert,
\end{equation}
with the $m$-photon energy shift
\begin{equation}
\delta^{[m]}=\frac{g_{0}^{2}m^{2}}{\omega_{M}-mg_{\text{cK}}}.\label{deltamdef}
\end{equation}
The eigensystem of the Hamiltonian $\hat{\tilde{H}}_{\text{gom}}$ can be expressed as
\begin{eqnarray}
\hat{\tilde{H}}_{\text{gom}}\vert m\rangle_{a}\vert n\rangle_{b}=E_{m,n}\vert m\rangle_{a}\vert n\rangle_{b}, \label{tildeH}
\end{eqnarray}
where $\vert n\rangle_{b}$ ($n= 0,1,2,\cdots$) are number states of the mechanical mode. The corresponding eigenvalues are
\begin{equation}
E_{m,n}=m\omega_{c}+(\omega_{M}-mg_{\text{cK}})n-\delta^{[m]}.\label{eigenenergyH}
\end{equation}

The eigensystem of the Hamiltonian $\hat{H}_{\text{gom}}$ can be obtained as
\begin{eqnarray}
\hat{H}_{\text{gom}}\vert m\rangle_{a}\vert\tilde{n}(m)\rangle_{b}=E_{m,n}\vert m\rangle_{a}\vert\tilde{n}(m)\rangle_{b},
\end{eqnarray}
where we introduce the $m$-photon displaced number states of the mechanical mode as
\begin{equation}
\vert\tilde{n}(m)\rangle_{b}\equiv\exp[\xi^{[m]}(\hat{b}^{\dagger}-\hat{b})]\vert n\rangle_{b}.
\end{equation}

The eigenstates of the Hamiltonian $\hat{H}_{\text{gom}}$ are direct product states of the photon number state $\vert m\rangle_{a}$ for mode $\hat{a}$ and the photon-number-dependent displaced number state $\vert\tilde{n}(m)\rangle_{b}$ for mode $\hat{b}$. For a given photon state $\vert m\rangle_{a}$, the $m$-photon displaced number states for mode $\hat{b}$ form a complete set of basis in the Hilbert space of the mechanical mode: $\sum_{n=0}^{\infty}\vert\tilde{n}(m)\rangle_{bb}\langle\tilde{n}(m)\vert=I_{b}$, where $I_{b}$ is the identity operator for mode $\hat{b}$. For studying few-photon optomechanical effects, we show the eigenenergy levels of $\hat{H}_\text{gom}$ in the subspace associated with zero, one, and two photons in Fig.~\ref{Fig1}(b). Physically, the induced optical nonlinearity depicted by $\delta^{[m]}$ [cf. $\delta^{[1]}$ and $\delta^{[2]}$ in Fig.~\ref{Fig1}(b)] is the origin of the photon blockade effect in this generalized optomechanical model. In the absence of the cross-Kerr interaction, i.e., $g_{\text{cK}}=0$, this optical nonlinearity becomes the Kerr nonlinearity in a standard optomechanical model~\cite{Rabl2011PRL, Liao2013PRA}. In addition, the photon-number-dependent displacement $\xi^{[m]}$ in this model is not a linear function of the photon number $m$. This nonlinear conditional photon displacement can be used to create quantum superposition states of the mechanical mode. When $g_{\text{cK}}=0$, the photon-number-dependent displacement $\xi^{[m]}$ is reduced to $mg_{0}/\omega_{M}$, which is the $m$-photon-induced mechanical displacement in the case of a typical optomechanical model~\cite{Marshall2003PRL}.

\section{Photon blockade effect \label{photonblockade}}

In this section, we study the photon blockade effect in the generalized optomechanical system by seeking the approximate analytical results and the exact numerical results.

\subsection{Analytical results}

To show the photon blockade effect, we introduce a monochromatic driving field to the cavity. The driving Hamiltonian is given by
\begin{equation}
\hat{H}_{d}=\Omega(\hat{a}^{\dagger}e^{i\omega_{d} t}+\hat{a}e^{-i\omega_{d} t}),\label{Hdri}
\end{equation}
where $\Omega$ and $\omega_{d}$ are the driving strength and driving frequency, respectively. Then the total Hamiltonian of the system becomes
\begin{equation}
\hat{H}_{\text{sys}}=\hat{H}_{\text{gom}}+\hat{H}_{d}.
\end{equation}
For below convenience, we work in a frame rotating at the driving frequency $\omega_{d}$, then the Hamiltonian of the total system becomes
\begin{equation}
\hat{H}_{\text{sys}}^{(I)}=\hat{H}_{\text{gom}}^{(I)}+\Omega(\hat{a}^{\dagger}+\hat{a}),\label{HsysInt}
\end{equation}
with
\begin{equation}
\hat{H}_{\text{gom}}^{(I)}=\Delta_{c}\hat{a}^{\dagger}\hat{a}+\omega_{M}\hat{b}^{\dagger}\hat{b}-g_{0}\hat{a}^{\dagger}\hat{a}(\hat{b}^{\dagger}+\hat{b})-g_{\text{cK}}\hat{a}^{\dagger}\hat{a}\hat{b}^{\dagger}\hat{b},
\end{equation}
where $\Delta_{c}=\omega_{c}-\omega_{d}$ is the detuning of the cavity frequency with respect to the driving frequency.
The eigensystem of $\hat{H}_{\text{gom}}^{(I)}$ can be written as
\begin{equation}
\hat{H}_{\text{gom}}^{(I)}\vert m\rangle_{a}\vert\tilde{n}(m)\rangle_{b}=\varepsilon_{m,n}\vert m\rangle_{a}\vert\tilde{n}(m)\rangle_{b},
\end{equation}
where the eigenvalue is defined by
\begin{eqnarray}
\varepsilon_{m,n}&=&E_{m,n}-m\omega_{d}\nonumber\\
&=&m\Delta_{c}+(\omega_{M}-mg_{\text{cK}})n-\delta^{[m]},
\end{eqnarray}
where $\delta^{[m]}$ is given by Eq.~(\ref{deltamdef}).

To analyze the photon blockade effect in the cavity, we analytically calculate the equal-time second-order correlation function of the cavity photons. To include the influence of the photon dissipation on the photon blockade, we phenomenologically add a non-Hermitian term to Hamiltonian~(\ref{HsysInt}) as follows
\begin{equation}
\hat{H}_{\text{eff}}=\hat{H}_{\text{sys}}^{(I)}- i\frac{\kappa}{2}\hat{a}^{\dagger}\hat{a},
\end{equation}
where $\kappa$ is the decay rate of the cavity field. In the following analytical calculations, we only consider the dissipation of the cavity mode because the optical dissipation dominates the dissipations of the system. However, the mechanical dissipation will be included in our numerical calculations.

In the weak-driving regime ($\Omega\ll\kappa$), the cavity is excited weakly and the average photon number in the cavity is small, then we can restrict the cavity field within the few-photon subspace spanned by these basis states $\{\vert 0\rangle_{a},\vert 1\rangle_{a},\vert 2\rangle_{a}\}$. In this subspace, a general state of the system can be written as
\begin{eqnarray}
\vert\varphi(t)\rangle&=&\sum\limits_{n=0}^{\infty}C_{0,n}(t)\vert 0\rangle_{a}\vert n\rangle_{b} +\sum\limits_{n=0}^{\infty}C_{1,n}(t)\vert
1\rangle_{a}\vert\tilde{n}(1)\rangle_{b} \notag \\
&&+\sum\limits_{n=0}^{\infty}C_{2,n}(t)\vert 2\rangle_{a}\vert\tilde{n}(2)\rangle_{b},
\end{eqnarray}
where $C_{0,n}(t)$, $C_{1,n}(t)$, and $C_{2,n}(t)$ are the probability amplitudes corresponding to the basis states $\vert0\rangle_{a}\vert n\rangle_{b}$, $\vert1\rangle_{a}\vert\tilde{n}(1)\rangle_{b}$, and $\vert 2\rangle_{a}\vert\tilde{n}(2)\rangle_{b}$, respectively. Based on the Schr\"{o}dinger equation $i\vert\dot{\varphi}(t)\rangle=\hat{H}_{\text{eff}}\vert\varphi(t)\rangle$, the equations of motion for these probability amplitudes can be obtained as
\begin{subequations}
\label{proampeq}
\begin{align}
\dot{C}_{0,n}=&-i\varepsilon_{0,n}C_{0,n}-i\Omega\sum\limits_{l=0}^{\infty}\;_{b}\langle n\vert\tilde{l}(1)\rangle_{b}C_{1,l},\\
\dot{C}_{1,n}=&-(i\varepsilon_{1,n}+\kappa/2)C_{1,n}-i\Omega\sum\limits_{l=0}^{\infty}\;_{b}\langle\tilde{n}(1)\vert l\rangle_{b}C_{0,l}\nonumber\\
&-i\sqrt{2}\Omega\sum\limits_{l=0}^{\infty}\;_{b}\langle\tilde{n}(1)\vert\tilde{l}(2)\rangle_{b}C_{2,l},\\
\dot{C}_{2,n}=&-(i\varepsilon_{2,n}+\kappa)C_{2,n}-i\sqrt{2}\Omega\sum\limits_{l=0}^{\infty}\;_{b}\langle\tilde{n}(2)\vert\tilde{l}(1)\rangle_{b}C_{1,l}.
\end{align}
\end{subequations}

In this system, the optical driving will induce the transitions among the states corresponding to neighboring photon numbers [i.e., the states in neighboring potential wells in Fig.~\ref{Fig1}(b)].
The magnitudes of these transitions are determined by the driving amplitude $\Omega$ and these Franck-Condon factors which are the inner products between these displaced number states in neighboring potential wells. This is because the optical driving induces photon hopping one by one. The values of these Franck-Condon factors can be calculated by the relation $\;_{b}\langle \tilde{n}(m)\vert\tilde{l}(m^\prime)\rangle_{b}=\;_{b}\langle
n\vert \exp[({\xi^{[m^\prime]}-\xi^{[m]})(\hat{b}^{\dagger}-\hat{b})}]\vert l\rangle_{b} $ $(m,m^\prime=0,1,2)$.
Here the matrix elements of the displacement operator in the Fock-state space can be calculated using the following relation~\cite{Oliveira1990PRA}
\begin{align}
\label{displaopr}
\;_{b}\langle n\vert D_{b}(x)\vert l\rangle_{b}=\left\{\begin{aligned}
\sqrt{\frac{n!}{l!}}e^{-\frac{\vert x\vert^{2}}{2}}(-x^{\ast})^{l-n}L_{n}^{l-n}(\vert x\vert^{2}),l\geq n \\
\sqrt{\frac{l!}{n!}}e^{-\frac{\vert x\vert^{2}}{2}}(x)^{n-l}L_{l}^{n-l}(\vert x\vert^{2}),n>l,
\end{aligned}\right.
\end{align}
where $\hat{D}_{b}(x)=\exp(x \hat{b}^\dagger-x^{\ast}\hat{b})$ is a displacement operator and $L_{n}^{l}(x)$ are the associated Laguerre polynomials.

In the weak-driving case, Eq.~(\ref{proampeq}) can be solved approximately by using a perturbation method, namely discarding the higher-order terms in the equations of motion for the lower-order variables.
We consider the case where initially the cavity is empty, namely $C_{1,n}(0)=0$ and $C_{2,n}(0)=0$, then the long-time solution of Eq.~(\ref{proampeq}) can be approximately obtained as
\begin{subequations}
\begin{align}
C_{0,n}(\infty)=&C_{0,n}(0)e^{-i\varepsilon_{0,n}t}, \\
C_{1,n}(\infty)=&-\Omega\sum\limits_{l=0}^{\infty}\frac{\;_{b}\langle\tilde{n}(1)\vert l\rangle_{b}C_{0,l}(0)e^{-i\varepsilon_{0,l}t}}{\varepsilon_{1,n}-\varepsilon_{0,l}-i\kappa/2}, \\
C_{2,n}(\infty)=&\sqrt{2}\Omega^{2}\sum\limits_{l,m=0}^{\infty}\frac{_{b}\langle \tilde{n}(2)\vert\tilde{l}(1)\rangle_{b}\;_{b}\langle\tilde{l}(1)\vert m\rangle_{b}}{(\varepsilon_{2,n}-\varepsilon_{0,m}-i\kappa)}\notag \\
& \times\frac{C_{0,m}(0)e^{-i\varepsilon_{0,m}t}}{(\varepsilon_{1,l}-\varepsilon_{0,m}-i\kappa/2)}.
\end{align}
\end{subequations}
Therefore, the normalized state of the system can be expressed as
\begin{eqnarray}
\vert\psi(t)\rangle&=&\mathcal{N}\sum\limits_{n=0}^{\infty}\left[C_{0,n}(t)\vert
0\rangle_{a}\vert n\rangle_{b}+C_{1,n}(t)\vert
1\rangle_{a}\vert\tilde{n}(1)\rangle_{b}\right.\notag \\
&&\left.+C_{2,n}(t)\vert 2\rangle_{a}\vert\tilde{n}(2)\rangle_{b}\right],
\end{eqnarray}
where we introduce a new normalization constant
\begin{equation}
\mathcal{N}=\left(\sum\limits_{s=0,1,2}\sum\limits_{n=0}^{\infty}\vert C_{s,n}(t)\vert^{2}\right)^{-1/2}.
\end{equation}
In the following calculations, we omit this normalization constant because of $\mathcal{N}\approx1$ in the weak-driving case.

The equal-time second-order correlation function in the weak-driving case can be written as
\begin{eqnarray}
g^{(2)}(0)\equiv\frac{\langle \hat{a}^{\dagger}\hat{a}^{\dagger}\hat{a}\hat{a} \rangle}{\langle \hat{a}^{\dagger}\hat{a} \rangle^2}=\frac{2P_{2}}{(P_{1}+2P_{2})^2}\approx\frac{2P_{2}}{P_{1}^2}, \label{analyticalg2}
\end{eqnarray}
where the photon probabilities are given by
\begin{eqnarray}
P_{m=0,1,2}=\sum\limits_{n=0}^{\infty}\vert C_{m,n}\vert^{2}.
\end{eqnarray}
We consider the case where the initial state of the mechanical resonator is $\vert 0\rangle_{b}$, i.e., $C_{0,n}(0)=\delta_{0,n}$, then the single- and two-photon probabilities in the long-time limit can be obtained as
\begin{subequations}
\label{PhoProb12}
\begin{align}
P_{1}=&\sum\limits_{n=0}^{\infty}\left\vert\frac{\Omega\;_{b}\langle\tilde{n}(1)\vert 0\rangle_{b}}{\Delta_{c}+n(\omega_{M}-g_{\text{cK}})-\delta^{[1]}-i\kappa/2}\right\vert^{2}.\\
P_{2}=&\sum\limits_{n=0}^{\infty}\left\vert\sum\limits_{l=0}^{\infty}\frac{\sqrt{2}\Omega^{2}\;_{b}\langle\tilde{n}(2)\vert\tilde{l}(1)\rangle_{b}}{[2\Delta_{c}+n(\omega_{M}-2g_{\text{cK}})-\delta^{[2]}-i\kappa]}\right. \notag \\
& \left.\times\frac{\;_{b}\langle\tilde{l}(1) \vert 0\rangle_{b}}{[\Delta_{c}+l(\omega_{M}-g_{\text{cK}})-\delta^{[1]}-i\kappa/2]}\right\vert^{2}.
\end{align}
\end{subequations}
By substituting Eq.~(\ref{PhoProb12}) into Eq.~(\ref{analyticalg2}), we can obtain the analytical result of the equal-time second-order correlation function $g^{(2)}(0)$. Note that this result is valid even in the parameter space $g_{0}\gg\omega_{M}$, under which the high-order phonon sidebands are still important. A concise result can be obtained in the Lamb-Dicke-like regime in which the first-order sideband is considered. When $\xi^{[s]} \ll 1$ $(s=1,2)$, we expand the matrix elements of the displacement operator to the first order of $\xi^{[s]}$, then the Franck-Condon factors can be approximated as
\begin{eqnarray}
\;_{b}\langle \tilde{n}(m)\vert\tilde{l}(m^\prime)\rangle_{b}&\approx& \delta_{n,l}-(\xi^{[m]}-\xi^{[m^\prime]})\sqrt{l+1}\delta_{n,l+1} \notag \\
&&+(\xi^{[m]}-\xi^{[m^\prime]})\sqrt{l}\delta_{n,l-1}.
\end{eqnarray}
Accordingly, the approximate expressions of the photon probabilities can be obtained as
\begin{eqnarray}
P_{1}&\approx&\frac{\Omega^{2}}{(\Delta_{c}-\delta^{[1]})^{2}+\kappa^{2}/4}, \notag \\
P_{2}&\approx&\frac{2\Omega^{4}}{[(2\Delta_{c}-\delta^{[2]})^{2}+\kappa^{2}][(\Delta_{c}-\delta^{[1]})^{2}+\kappa^{2}/4]}.
\end{eqnarray}

In this case, the second-order correlation function takes the form
\begin{eqnarray}
g^{(2)}(0)&\approx&\frac{2P_{2}}{P_{1}^{2}}=\frac{4(\Delta_{c}-\delta^{[1]})^{2}+\kappa^{2}}{(2\Delta_{c}-\delta^{[2]})^{2}+\kappa^{2}}.
\end{eqnarray}
Based on these analytical results, we can obtain the optimal driving frequencies corresponding to the single- and two-photon resonance processes.

In the single-photon resonance (spr) case, $\Delta_{c}=\delta^{[1]}$, the correlation function becomes
\begin{equation}
g^{(2)}_{\text{spr}}(0)\approx \frac{\kappa^{2}}{(2\delta^{[1]}-\delta^{[2]})^{2}+\kappa^{2}}.
\end{equation}
Owing to $2\delta^{[1]}\neq\delta^{[2]}$, we have $g^{(2)}_{\text{spr}}(0)<1$, which corresponds to sub-Poisson distribution of photons. In particular, when $\delta^{[2]}-2\delta^{[1]}\gg\kappa$, the two-photon probability is largely suppressed, then the photon blockade effect takes place in this generalized optomechanical system.

In the two-photon resonance (tpr) case, $\Delta_{c}=\delta^{[2]}/2$, the correlation function is reduced to
\begin{equation}
g^{(2)}_{\text{tpr}}(0)\approx\frac{(\delta^{[2]}-2\delta^{[1]})^{2}+\kappa^{2}}{\kappa^{2}}.
\end{equation}
Here, the second-order correlation function could be much larger than $1$, and then we can observe photon-assisted tunneling in this system~\cite{Liu2013PRA}.

\subsection{Numerical results}

To include the dissipations of the cavity field and the mechanical resonator, in this section we study the photon blockade effect in the open-system case by using the method of quantum master equation. In particular, we assume that the cavity field and the mechanical resonator are connected with a vacuum bath and a heat bath at temperature $T$, respectively. Under the Born-Markov approximation and the rotating-wave approximation, the quantum master equation in the rotating frame is written as
\begin{eqnarray}
\frac{d\hat{\rho}(t)}{dt}&=&-i[\hat{H}_{\text{sys}}^{(I)},\hat{\rho}(t)]+\kappa \mathcal{D}[\hat{a}]\hat{\rho}(t) +\gamma_{M}(\bar{n}_{M}+1) \mathcal{D}[\hat{b}] \hat{\rho}(t) \notag \\
&&+\gamma_{M}\bar{n}_{M}\mathcal{D}[\hat{b}^{\dagger}] \hat{\rho}(t),\label{SME}
\end{eqnarray}
where $\kappa$ and $\gamma_{M}$ are, respectively, the decay rates of the cavity field and the mechanical oscillator. The $\bar{n}_{M}=(e^{\hbar\omega_{M}/(k_{B}T)}-1)^{-1}$ is the average thermal phonon number associated with the mechanical dissipation, with $k_{B}$ being the Boltzmann constant. The Lindblad superoperators used in Eq.~(\ref{SME}) are defined by
\begin{equation}
\mathcal{D}[\hat{o}]\hat{\rho}(t)=\frac{1}{2}[2\hat{o}\hat{\rho}(t)\hat{o}^{\dagger}-\hat{o}^{\dagger}\hat{o}\hat{\rho}(t)-\hat{\rho}(t)\hat{o}^{\dagger}\hat{o}]
\end{equation}
with $\hat{o}=\hat{a}$, $\hat{b}$, and $\hat{b}^{\dag}$. The three Lindblad superoperators $\mathcal{D}[\hat{a}]\hat{\rho}(t)$, $\mathcal{D}[\hat{b}]\hat{\rho}(t)$, and $\mathcal{D}[\hat{b}^{\dagger }]\hat{\rho}(t)$ in Eq.~(\ref{SME}) describe the cavity-field dissipation, the mechanical damping, and the mechanical thermal excitation, respectively.

By numerically solving Eq.~(\ref{SME}), we can get the steady-state density operator $\hat{\rho}_{\text{ss}}$ of the system, and then the photon-number probabilities $P_{m=0,1,2}=\text{Tr}[\sum\limits_{n=0}^{\infty}\vert m \rangle_a \vert n\rangle_b\;_a\!\langle m \vert _b\langle n \vert \hat{\rho}_{\text{ss}}]$ can be calculated numerically. The second-order correlation function $g^{(2)}(0)$ can also be obtained by $g^{(2)}(0)=\text{Tr}(\hat{a}^{\dagger}\hat{a}^{\dagger}\hat{a}\hat{a}\hat{\rho}_{\text{ss}})/[\text{Tr}(\hat{a}^{\dagger}\hat{a}\hat{\rho}_{\text{ss}})]^2$.

\begin{widetext}
\begin{center}
\begin{table}[ht!]
\caption{The correspondence among the single- and two-photon transitions $\vert 0\rangle_{a}\vert 0\rangle_{b}\rightarrow \vert 1\rangle_{a}\vert\tilde{n}(1)\rangle_{b}$ and $\vert 0\rangle_{a}\vert 0\rangle_{b}\rightarrow \vert 2\rangle_{a}\vert\tilde{n}(2)\rangle_{b}$, the values of the optimal driving detuning $\Delta_{c}/\omega_{M}$ related to these resonant transitions, and the marks of these peaks and dips in the second-order correlation function $g^{(2)}(0)$ [cf. Fig.~\ref{Fig2}(b)].}
\label{Tab}
\begin{tabular}{|c|c|c|c|c|c|c|c|}
\hline
\multicolumn{1}{|c|}{\multirow{3}{*}{\tabincell{c}{Single-photon resonant transitions \\ $\vert 0\rangle_{a}\vert 0\rangle_{b}\rightarrow \vert 1\rangle_{a}\vert\tilde{n}(1)\rangle_{b}$}}}& $\Delta_{c}/\omega_{M}$ & $0.594$ & $-0.231$ & $-1.056$ & $-1.881$ & $-2.706$ & $-3.531$ \\ \cline{2-8}
\multicolumn{1}{|c|}{\multirow{3}{*}{}}& Transition final states & $\vert 1\rangle_{a}\vert\tilde{0}(1)\rangle_{b}$ & $\vert 1\rangle_{a}\vert\tilde{1}(1)\rangle_{b}$ & $\vert 1\rangle_{a}\vert\tilde{2}(1)\rangle_{b}$ & $\vert 1\rangle_{a}\vert\tilde{3}(1)\rangle_{b}$ & $\vert 1\rangle_{a}\vert\tilde{4}(1)\rangle_{b}$ & $\vert 1\rangle_{a}\vert\tilde{5}(1)\rangle_{b}$ \\ \cline{2-8}
\multicolumn{1}{|c|}{\multirow{3}{*}{}}& Marks & $d_{0}$ & $d_{1}$ & $d_{2}$ & $d_{3}$ & $d_{4}$ & $d_{5}$ \\ \hline
\multicolumn{1}{|c|}{\multirow{3}{*}{\tabincell{c}{Two-photon resonant transitions \\ $\vert 0\rangle_{a}\vert 0\rangle_{b}\rightarrow \vert 2\rangle_{a}\vert\tilde{n}(2)\rangle_{b}$}}} & $\Delta_{c}/\omega_{M}$ & $1.508$ & $1.183$ & $0.858$ & $0.533$ & $-0.117$ & $-1.092$ \\ \cline{2-8}
\multicolumn{1}{|c|}{\multirow{3}{*}{}}&Transition final states & $\vert 2\rangle_{a}\vert\tilde{0}(2)\rangle_{b}$ & $\vert 2\rangle_{a}\vert\tilde{1}(2)\rangle_{b}$ & $\vert 2\rangle_{a}\vert\tilde{2}(2)\rangle_{b}$ & $\vert 2\rangle_{a}\vert\tilde{3}(2)\rangle_{b}$ & $\vert 2\rangle_{a}\vert\tilde{5}(2)\rangle_{b}$ & $\vert 2\rangle_{a}\vert\tilde{8}(2)\rangle_{b}$ \\ \cline{2-8}
\multicolumn{1}{|c|}{\multirow{3}{*}{}}& Marks & $p_{0}$ & $p_{1}$ & $p_{2}$ & $p_{3}$ & $p_{5}$ & $p_{8}$ \\ \hline
\end{tabular}
\end{table}
\end{center}
\end{widetext}

\begin{figure}[tbp]
\center
\includegraphics[bb=0 0 279 327, width=0.47 \textwidth]{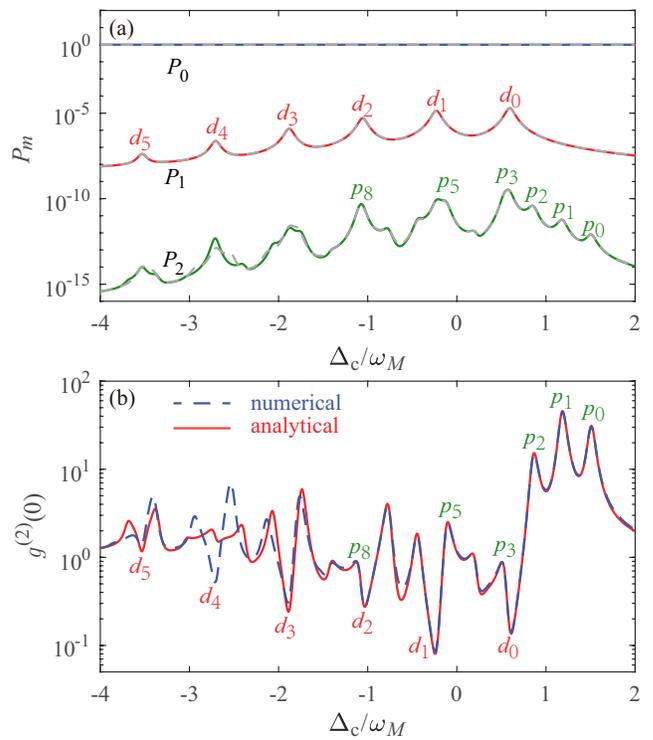}
\caption{(Color online) (a) The photon-number probabilities $P_{m=0,1,2}$ as a function of the driving detuning $\Delta_{c}/\omega_{M}$. The colored solid curves and the gray dashed curves are plotted based on the analytical and numerical results, respectively. (b) The equal-time second-order correlation function $g^{(2)}(0)$ as a function of the driving detuning $\Delta_{c}/\omega_{M}$. The red solid curve and the blue dashed curve correspond to the analytical and numerical results, respectively. Other parameters are given by $g_{0}/\omega_{M}=0.7$, $g_{\text{cK}}/g_{0}=0.25$, $\kappa/\omega_{M}=0.1$, $\gamma_{M}/\omega_{M}=0.001$, $\Omega/\kappa=0.01$, and $\bar{n}_{M}=0$.}
\label{Fig2}
\end{figure}

To seek for an optimal driving detuning of the photon blockade, we investigate the dependence of the cavity photon-number distributions on the driving detuning. In Fig.~\ref{Fig2}(a), we plot both the analytical results (the colored solid curves) and the numerical results (the gray dashed curves) of the photon-number probabilities $P_{m=0,1,2}$ as a function of the driving detuning $\Delta_{c}/\omega_{M}$. Here we can see the relations $P_{0}\approx1$ and $P_{0}\gg P_{1}\gg P_{2}$ in the weak-driving case. In addition, Fig.~\ref{Fig2}(a) shows that there are some peaks in the single-photon probability (the red solid curve) and the two-photon probability (the green solid curve). By analyzing the single- and two-photon resonance conditions, we find that the locations of these peaks in the curves of $P_{1}$ and $P_{2}$ are determined by the single- and two-photon resonant transitions $\vert 0\rangle_{a}\vert 0\rangle_{b}\rightarrow \vert 1\rangle_{a}\vert\tilde{n}(1)\rangle_{b}$ and $\vert 0\rangle_{a}\vert 0\rangle_{b}\rightarrow \vert 2\rangle_{a}\vert\tilde{n}(2)\rangle_{b}$, respectively. To be more clear, we mark these peaks in the single- and two-photon probabilities $P_{1}$ and $P_{2}$ as $d_{n}$ and $p_{n}$, respectively. The subscripts in $d_{n}$ and $p_{n}$ correspond to the quantum numbers in the states $\vert 1\rangle_{a}\vert\tilde{n}(1)\rangle_{b}$ and $\vert 2\rangle_{a}\vert\tilde{n}(2)\rangle_{b}$ involved in these transitions. It follows from the relation $g^{(2)}(0)\approx 2P_{2}/P_{1}^2$ that the peak values of $P_{2}$ and $P_{1}$ are related to the peaks and dips in $g^{(2)}(0)$, respectively. That is why we mark the peaks in $P_{1}$ as $d_{n}$. In Table~\ref{Tab}, we present the correspondence among these transitions $\vert 0\rangle_{a}\vert 0\rangle_{b}\rightarrow \vert 1\rangle_{a}\vert\tilde{n}(1)\rangle_{b}$ and $\vert 0\rangle_{a}\vert 0\rangle_{b}\rightarrow \vert 2\rangle_{a}\vert\tilde{n}(2)\rangle_{b}$, the locations (i.e., the values of $\Delta_{c}/\omega_{M}$) of these peaks and dips in $g^{(2)}(0)$, and the marks of these peaks and dips.

The dependence of the photon blockade effect on the driving detuning can be analyzed by plotting the correlation function $g^{(2)}(0)$ as a function of $\Delta_{c}/\omega_{M}$. In Fig.~\ref{Fig2}(b), the red solid curve is plotted based on the analytical results given in Eq.~(\ref{analyticalg2}), while the blue dashed curve is plotted using the numerical solution of quantum master equation~(\ref{SME}). We can see that the analytical results can match well with the numerical results. By comparing the correlation function $g^{(2)}(0)$ with the photon number probabilities $P_{m=0,1,2}$, we see that the locations of these dips and peaks of $g^{(2)}(0)$ correspond to single- and two-photon resonant transitions, respectively. Figure~\ref{Fig2}(b) also shows that the photon blockade effect [$g^{(2)}(0)\ll1$, corresponding to the dips in the correlation function] can be observed at the single-photon resonance.
\begin{figure}[tbp]
\center
\includegraphics[bb=0 0 279 327, width=0.47 \textwidth]{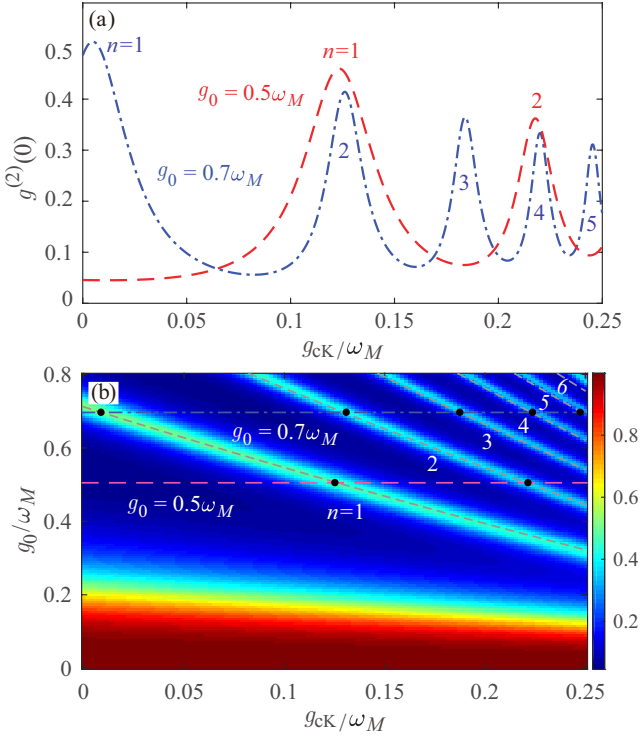}
\caption{(Color online) (a) Plot of $g^{(2)}(0)$ as a function of $g_{\text{cK}}/\omega_{M}$ under $g_{0}/\omega_{M}=0.5$, $0.7$ and the single-photon resonance $\Delta_c=\delta^{[1]}$. (b) Plot of $g^{(2)}(0)$ as a function of $g_{\text{cK}}/\omega_{M}$ and $g_{0}/\omega_{M}$ at $\Delta_{c}=\delta^{[1]}$. The grey dashed curves refer to Eq.~(\ref{simresoeq}). Other parameters are $\kappa/\omega_{M}=0.1$, $\gamma_M/\omega_{M}
=0.001$, $\Omega/\kappa=0.01$, and $\bar{n}_{M}=0$.}
\label{Fig3}
\end{figure}

We proceed to study the influence of the cross-Kerr interaction on the photon blockade. In Fig.~\ref{Fig3}(a) we plot the equal-time second-order correlation function $g^{(2)}(0)$ at the steady state as a function of the cross-Kerr parameter $g_{\text{cK}}/\omega_{M}$, under given values of $g_{0}/\omega_{M}$ and single-photon resonant driving $\Delta_{c}=\delta^{[1]}$. Here we can see that the correlation function exhibits an oscillating pattern with several resonance peaks located at specific values of $g_{\text{cK}}/\omega_{M}$. Moreover, the cross-Kerr interaction could either enhance or suppress the photon blockade effect, as shown in the cases corresponding to $g_{0}/\omega_{M}=0.7$ and $g_{0}/\omega_{M}=0.5$. A more comprehensive analysis of these phenomena is shown in Fig.~\ref{Fig3}(b), in which we plot the correlation function as a function of $g_{0}/\omega_{M}$ and $g_{\text{cK}}/\omega_{M}$ under the single-photon resonant transition $\vert 0\rangle_{a}\vert 0\rangle_{b}\rightarrow \vert 1\rangle_{a}\vert\tilde{0}(1)\rangle_{b}$. The results display that the value of $g^{(2)}(0)$ is around $1$ when $g_{0}/\omega_{M}<0.2$. For a given value of $g_{\text{cK}}$, the correlation function $g^{(2)}(0)$ experiences some oscillations with the increasing of the ratio $g_{0}/\omega_{M}$. The locations of these resonant peaks are determined by the two-photon resonant transitions involving these phonon sidebands ($\vert 1\rangle_{a}\vert\tilde{0}(1)\rangle_{b}\rightarrow\vert 2\rangle_{a}\vert\tilde{n}(2)\rangle_{b}$), and hence the resonant peaks are related to the corresponding phonon sideband indexes $n$. When the cross-Kerr parameter $g_{\text{cK}}/\omega_{M}$ changes, these joined resonant peaks form resonant curves in the $2$D plot, as marked by the grey dashed curves in Fig.~\ref{Fig3}(b). In this optomechanical system, there are many phonon sidebands and these phonon sideband channels could induce single- and two-photon resonant transitions simultaneously. Therefore, the locations of these resonant peaks are determined by the single- and two-photon resonant transitions, which depend on the optomechanical coupling strength $g_{0}/\omega_{M}$ and the cross-Kerr parameter $g_{\text{cK}}/\omega_{M}$. By analyzing the single- and two-photon resonant transitions, the parameter equation determining the locations of these resonant curves can be obtained as
\begin{equation}
\frac{g_0}{\omega_M}=\left[\frac{n}{2}\left(1-\frac{2g_{\text{cK}}}{\omega_M}\right)^{2}\left(1-\frac{g_{\text{cK}}}{\omega_M}\right)\right]^{1/2},\hspace{0.2 cm}n=1,2,\cdots. \label{simresoeq}
\end{equation}
In particular, when the cross-Kerr interaction is absence, i.e., $g_{\text{cK}}/\omega_{M}=0$, Eq.~(\ref{simresoeq}) is reduced to $g_0/\omega_M=\sqrt{n/2}$, which comes back to the result obtained for the typical optomechanical model~\cite{Liao2013PRA}. Note that the locations of these peaks in Fig.~\ref{Fig3}(a) correspond to these black spots crossed by the lines at $g_{0}/\omega_{M}=0.5$, $0.7$ and these resonant curves in Fig.~\ref{Fig3}(b).
\begin{figure}[tbp]
\center
\includegraphics[bb=0 0 279 327, width=0.47 \textwidth]{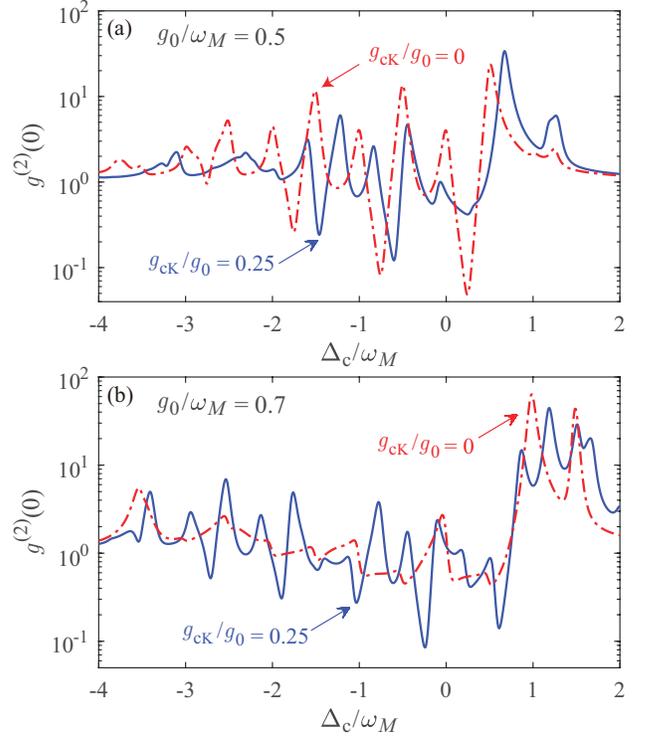}
\caption{(Color online) Plot of $g^{(2)}(0)$ as a function of the driving detuning $\Delta_{c}/\omega_{M}$ when (a) $g_{0}/\omega_{M}=0.5$ and (b) $g_{0}/\omega_{M}=0.7$. The red dot-dashed and blue solid curves correspond to the cases of $g_{\text{cK}}/g_0=0$ and $g_{\text{cK}}/g_0=0.25$, respectively. Other parameters are given by $\kappa/\omega_{M}=0.1$, $\gamma_{M}/\omega_{M}=0.001$, $\Omega/\kappa=0.01$, and $\bar{n}_{M}=0$.}
\label{Fig4}
\end{figure}

To further illustrate the effect of the cross-Kerr interaction on the photon blockade at different values of $g_{0}/\omega_{M}$, we plot the steady-state correlation function $g^{(2)}(0)$ as a function of the driving detuning $\Delta_{c}/\omega_{M}$ at $g_0/\omega_{M}=0.5$ and $0.7$, as shown in Figs.~\ref{Fig4}(a) and~\ref{Fig4}(b), respectively. Here, the red dot-dashed and blue solid curves correspond to the two cases of $g_{\text{cK}}/g_0=0$ and $g_{\text{cK}}/g_0=0.25$, respectively. It can be confirmed that the optimal driving frequencies at these dips correspond to the single-photon resonant transitions $\vert 0\rangle_{a}\vert 0\rangle_{b}\rightarrow \vert 1\rangle_{a}\vert\tilde{n}(1)\rangle_{b}$. By comparing the dips in the two cases, we find that the optimal driving detuning of the photon blockade is changed due to the presence of the cross-Kerr interaction. The shift of the optimal driving frequency can be deduced from the eigenenergy spectrum~(\ref{eigenenergyH}) of the system. The influence of the cross-Kerr interaction in the cases of $g_0/\omega_{M}=0.5$ and $0.7$ is consistent with the results in Fig.~\ref{Fig3}.
\begin{figure}[tbp]
\center
\includegraphics[bb=0 0 279 327, width=0.47 \textwidth]{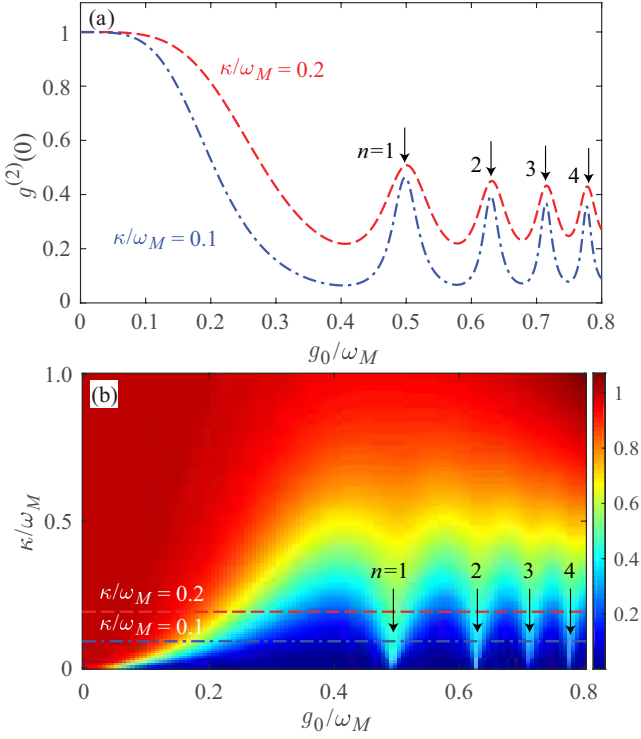}
\caption{(Color online) (a) Plot of $g^{(2)}(0)$ as a function of $g_{0}/\omega_{M}$ at different values of $\kappa/\omega_{M}$ when $\Delta_c=\delta^{[1]}$. (b) Plot of $g^{(2)}(0)$ as a function of $g_{0}/\omega_{M}$ and $\kappa/\omega_{M}$ at $\Delta_c=\delta^{[1]}$. Other parameters are $\gamma_{M}/\omega_{M}=0.001$, $g_{\text{cK}}/g_0=0.25$, $\Omega/\kappa=0.01$, and $\bar{n}_{M}=0$.}
\label{Fig5}
\end{figure}

As shown in the typical optomechanical model, the resolved-sideband condition should be satisfied to observe the photon blockade effect. Hence the decay rate of the cavity mode will significantly affect the photon blockade.
In Fig.~\ref{Fig5} we show the dependence of $g^{(2)}(0)$ on the optomechanical coupling strength $g_{0}/\omega_{M}$ under the single-photon resonant driving condition $\Delta_{c}=\delta^{[1]}$. In Fig.~\ref{Fig5}(a), the steady-state correlation function $g^{(2)}(0)$ is plotted as a function of $g_0/\omega_{M}$ at different values of $\kappa/\omega_{M}$ when $\Delta_{c}=\delta^{[1]}$. We can see that the correlation function exhibits an oscillating pattern with increasing $g_{0}/\omega_{M}$ owing to the modulation of the phonon sidebands. The locations of these resonant peaks correspond to the two-photon resonant transitions, which is consistent with the theoretical results given by Eq.~(\ref{simresoeq}). In addition, the photon blockade effect for the case of $\kappa/\omega_{M}=0.1$ is better than that for $\kappa/\omega_{M}=0.2$. To see a wider parameter space corresponding to the photon blockade, in Fig.~\ref{Fig5}(b), the steady-state correlation function $g^{(2)}(0)$ is plotted as a function of $g_{0}/\omega_{M}$ and $\kappa/\omega_{M}$ at $\Delta_{c}=\delta^{[1]}$. Here we see that the photon blockade effect ($g^{(2)}(0) \ll 1$) can be observed in the deep-resolved-sideband regime $\kappa/\omega_{M} < 0.1$.

\section{Generation of the Schr\"{o}dinger Cat states \label{Schcatstate}}

Another interesting topic in few-photon optomechanics is the generation of the Schr\"{o}dinger cat states in the mechanical resonator based on the conditional dynamics of the optomechanical coupling~\cite{Marshall2003PRL}. In this section, we will study the enhancement of the mechanical displacement induced by a single photon and the generation of macroscopic mechanical cat states. We will also study the Wigner function~\cite{Buzek} and the probability distribution of the rotated quadrature operator~\cite{Walls} in the generated cat states based on the analytical and numerical results.

\subsection{Analytical solution}

For the Hamiltonian $\hat{H}_{\text{gom}}$, its unitary evolution operator can be written as (see Appendix)
\begin{eqnarray}
\hat{U}(t)&=&e^{-i\omega_{c}t\hat{a}^{\dagger}\hat{a}}e^{i\hat{\mu}(t)
\hat{a}^{\dagger}\hat{a}\hat{a}^{\dagger}\hat{a}}e^{-i\hat{\nu}(t)\hat{a}^{\dagger}\hat{a}\hat{a}^{\dagger}\hat{a}\hat{a}^{\dagger}\hat{a}} \notag \\
&&\times e^{\hat{a}^{\dagger}\hat{a}[\hat{\lambda}(t)\hat{b}^{\dagger}-\hat{\lambda}^{\ast}(t)\hat{b}]}e^{i(g_{\text{cK}}\hat{a}^{\dagger}\hat{a}-
\omega_{M})t\hat{b}^{\dagger}\hat{b}}, \label{uniopr}
\end{eqnarray}
where we introduce the variables
\begin{subequations}
\label{uniopevariab}
\begin{align}
\hat{\mu}(t)=&\frac{g_{0}^{2}\{\omega_{M}t -\sin[(\omega_{M}-g_{\text{cK}}\hat{a}^{\dagger}\hat{a})t]\}}{(\omega_{M}-g_{\text{cK}}\hat{a}^{\dagger}\hat{a})^{2}}, \\
\hat{\nu}(t)=&\frac{g_{\text{cK}}g_{0}^{2}t}{(\omega_{M}-g_{\text{cK}}\hat{a}^{\dagger}\hat{a})^{2}}, \\
\hat{\lambda}(t)=&\frac{g_{0}}{\omega_{M}-g_{\text{cK}}\hat{a}^{\dagger}\hat{a}}[1-e^{i(g_{\text{cK}}\hat{a}^{\dagger}\hat{a}-\omega_{M})t}].
\end{align}
\end{subequations}
To generate the mechanical cat states, we consider an initial state $\vert\psi(0)\rangle=(\vert 0\rangle_{a}+\vert 1\rangle_{a})\vert 0\rangle_{b}/\sqrt{2}$ of the system,
where $\vert m \rangle_{a} (m=0,1)$ denotes the Fock state of the cavity field and $\vert 0 \rangle _{b}$ is the ground state of the mechanical resonator, which can be prepared via the
ground state cooling~\cite{Chan2011Nature,Teufel2011Nature}. By utilizing the unitary evolution operator $\hat{U}(t)$, the state of the system at time $t$ can be obtained as
\begin{eqnarray}
\vert\psi(t)\rangle &=&\hat{U}(t)\vert\psi(0)\rangle \notag \\
&=&\frac{1}{\sqrt{2}}[ \vert 0\rangle _{a}\vert0\rangle_{b}+e^{i\vartheta (t)}\vert 1\rangle_{a}\vert \beta(t)\rangle_{b}], \label{psitsta01basis}
\end{eqnarray}
where the phase factor $\vartheta(t)$ and the mechanical displacement $\beta(t)$ are defined by
\begin{subequations}
\label{varthetabeatt}
\begin{align}
\vartheta(t)=&-\omega_{c}t+\frac{g_{0}^{2}}{(\omega_{M}-g_{\text{cK}})^{2}}\{(\omega_{M}-g_{\text{cK}})t\notag\\
&-\sin[(\omega_{M}-g_{\text{cK}})t]\},\\
\beta(t)=&\frac{g_{0}}{\omega _{M}-g_{\text{cK}}}[1-e^{i(g_{\text{cK}}-\omega_{M}) t}].
\end{align}
\end{subequations}

By expanding the cavity-mode state with basis states $\vert\pm\rangle_{a}=(\vert 0\rangle_{a}\pm\vert 1\rangle_{a})/\sqrt{2}$, Eq.~(\ref{psitsta01basis}) becomes
\begin{eqnarray}
\vert\psi(t)\rangle &=&\frac{1}{2}\left[\frac{1}{\mathcal{N}_{+}(t)}\vert+\rangle_{a}\vert \Phi^{(+)}(t)\rangle_{b}+\frac{1}{\mathcal{N}_{-}(t)}\vert -\rangle _{a}\vert \Phi ^{(-)}(t)\rangle _{b}\right], \label{state}\nonumber\\
\end{eqnarray}
where we introduce the mechanical cat states
\begin{eqnarray}
\vert \Phi^{(\pm)}(t)\rangle_{b}=\mathcal{N}_{\pm}(t)[ \vert 0 \rangle_{b} \pm e^{i\vartheta (t)}\vert \beta (t) \rangle_{b}], \label{catstates}
\end{eqnarray}
which are quantum superposition of the ground state $\vert 0 \rangle_{b}$ and the coherent state $\vert\beta(t)\rangle_{b}$. The normalization constants $\mathcal{N}_{\pm}(t)$ are given by
\begin{equation}
\mathcal{N}_{\pm}(t)=\left[2\left(1\pm\cos[\vartheta (t)]e^{-\frac{\vert\beta (t)\vert^{2}}{2}}\right)\right]^{-1/2}.
\end{equation}
When we choose proper detection time $t=(2n+1)\pi/(g_{\text{cK}}-\omega_{M})$, a maximal value of $|\beta(t)|$ is obtained. To minimal the influence of environment noise, we choose $t_{s}=\pi/(g_{\text{cK}}-\omega_{M})$ as the detection time in the following discussions. Equation~(\ref{varthetabeatt}b) shows that the maximal displacement amplitude is $|\beta(t_{s})|=2g_{0}/(g_{\text{cK}}-\omega_{M})$.

When the cavity field is detected in states $\vert\pm\rangle_{a}$, the mechanical resonator will collapse into the mechanical cat states $\vert \Phi^{(\pm)}(t)\rangle_{b}$ accordingly. The corresponding detection probabilities are
\begin{equation}
\mathcal{P}_{\pm}(t)=\frac{1}{2}\left[1\pm\cos[\vartheta(t)] e^{-\frac{\vert\beta (t)\vert ^{2}}{2}}\right]=\frac{1}{4|\mathcal{N}_{\pm}(t)|^{2}}. \label{Meaprobabt}
\end{equation}
It can be seen from Eq.~(\ref{Meaprobabt}) that, for a sufficiently large displacement $\vert\beta(t)\vert$ such that $\exp[-\vert\beta(t)\vert^{2}/2]\approx0$, the detection probabilities become $\mathcal{P}_{\pm}(t)\approx1/2$.
\begin{figure}[tbp]
\center
\includegraphics[bb=2 10 271 341, width=0.47 \textwidth]{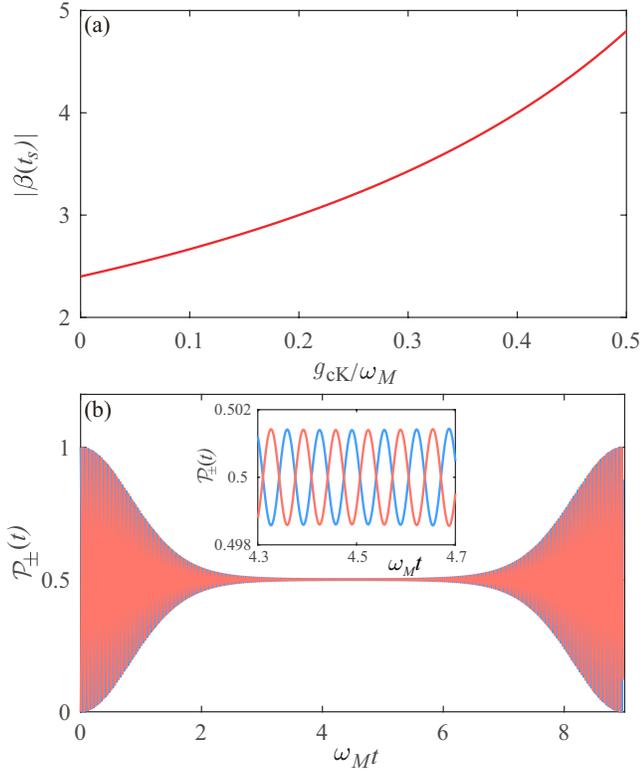}
\caption{(Color online) (a) Plot of $|\beta(t_s)|$ as a function of $g_{\text{cK}}/\omega_{M}$. (b) The probabilities $\mathcal{P}_{\pm}(t)$ as a function of time $t$.  The inset in (b) is a zoomed-in plot of $\mathcal{P}_{\pm}(t)$ as a function of time $t$. Other parameters are $\omega_{c}/\omega_{M}=100$ , $g_{0}/\omega_{M}=1.2$, and $g_{\text{cK}}/g_{0}=0.25$.}
\label{Fig6}
\end{figure}

In order to investigate how the maximal displacement amplitude $|\beta(t_{s})|$ depends on the cross-Kerr interaction $g_{\text{cK}}$, in Fig.~\ref{Fig6}(a), we plot $|\beta(t_{s})|$ as a function of $g_{\text{cK}}/\omega_{M}$. We can see that $|\beta(t_{s})|$ increases monotonically with the increase of the cross-Kerr interaction strength. Under the assistance of the cross-Kerr interaction, the cat states which are formed by quantum superposition of the vacuum state and coherent states can be obtained. To approximately distinguish the coherent state from the vacuum state, a coherent displacement $|\beta|>3$ is usually expected. Figure~\ref{Fig6}(a) shows that the superposition of distinct coherent states can be created under the cross-Kerr interaction. In Fig.~\ref{Fig6}(b), we display the time dependence of the probabilities $\mathcal{P}_{\pm}(t)$. Here we can see that the two probabilities $\mathcal{P}_{+}(t)$ and $\mathcal{P}_{-}(t)$ have similar envelops. In the intermediate duration around the detection time $t_{s}=\pi/(g_{\text{cK}}-\omega_{M})$, the oscillation amplitude is negligible [see the inset in panel (b)]. At the detection time $t_{s}\approx4.19$, the probabilities $\mathcal{P}_{+}(t_s)\approx\mathcal{P}_{-}(t_s)\approx1/2$, which is consist with the analysis based on the condition $\exp[-\vert\beta(t)\vert^{2}/2]\approx0$.

\subsection{The Wigner function and the probability distribution of the rotated quadrature operator}

The quantum interference and coherence effects in the generated mechanical cat states can be revealed by calculating either the Wigner function or the probability distribution of the rotated quadrature operator. For the mechanical mode in the density matrix $\hat{\rho}_b$, the Wigner function is defined by~\cite{Buzek}
\begin{equation}
W(\eta) =\frac{2}{\pi}\text{Tr}[\hat{\rho}_b \hat{D}(\eta)e^{i\pi \hat{b}^{\dagger}\hat{b}}\hat{D}^{\dagger}(\eta)],\label{Wignfundef}
\end{equation}
where $\hat{D}(\eta)=\exp(\eta\hat{b}^{\dagger}-\eta^{\ast}\hat{b})$ is a displacement operator. Corresponding to the
states $\vert\Phi^{(\pm)}(t)\rangle_{b}$ in Eq.~(\ref{catstates}), the Wigner functions can be obtained by substituting the density matrices $\hat{\rho}^{(\pm)}_{b}=\vert\Phi^{(\pm)}(t)\rangle_{bb}\langle\Phi^{(\pm)}(t)\vert$ into Eq.~(\ref{Wignfundef}) as
\begin{eqnarray}
W^{(\pm)}(\eta)&=&\frac{2\mathcal{N}_{\pm}^2}{\pi }\left(e^{-2\vert\eta\vert^{2}}+e^{-2\vert\beta (t)-\eta \vert^{2}}\right.\notag \\
&&\left.\pm 2\text{Re}[e^{-i\vartheta ( t)}e^{ -\frac{1}{2}\vert \beta (t) \vert ^{2}+2\beta^{\ast }(t)\eta -2\vert \eta \vert ^{2}}]\right).
\end{eqnarray}

For the rotated quadrature operator
\begin{equation}
\hat{X}({\theta})=\frac{1}{\sqrt{2}}(\hat{b}e^{-i\theta }+\hat{b}^{\dagger }e^{i\theta}),
\end{equation}
its eigenstate is denoted by $\vert X({\theta})\rangle_{b}$: $\hat{X}({\theta})\vert X({\theta})\rangle_{b}= X({\theta})\vert X({\theta})\rangle_{b}$~\cite{Walls}. For the states $\vert\Phi^{(\pm)}(t)\rangle_{b}$, we can obtain the probability distributions of the rotated quadrature operator $\hat{X}({\theta})$ as
\begin{eqnarray}
P^{(\pm)}\left[ X({\theta})\right] &=& \left\vert _{b}\langle X({\theta}) \vert \Phi ^{(\pm) }(t)\rangle _{b}\right\vert ^{2} \notag \\
&=&\mathcal{N}_{\pm}^2\left\vert _{b}\langle X({\theta}) \vert 0\rangle_{b}\pm e^{i\vartheta (t)}\;_{b}\langle X({\theta})\vert
\beta (t) \rangle_{b} \right\vert ^{2}. \notag \\
\end{eqnarray}
Here the inner product of the vacuum state $\vert 0\rangle _{b}$ and the coherent state $\vert \beta(t)\rangle_{b}$ with the eigenatate $\vert X({\theta})\rangle$ of the rotated quadrature operator can be calculated with the relations
\begin{subequations}
\begin{align}
_{b}\langle X({\theta}) \vert 0\rangle_{b}=&\frac{H_{0}[ X({\theta})]}{\sqrt{\pi^{1/2}}} e^{-X^{2}({\theta})/2}, \\
_{b}\langle X({\theta})\vert \beta (t)\rangle_{b}=&e^{-\vert \beta(t) \vert^{2}/2}\sum\limits_{n=0}^{\infty }\frac{[\beta (t)]^{n}H_{n}[X({\theta})]}{n!\sqrt{\pi^{1/2}2^{n}}} \notag \\
& \times e^{-X^{2}({\theta})/2}e^{-i\theta n},
\end{align}
\end{subequations}
where $H_{n}(x)$ are the Hermite polynomials.
\begin{figure}[tbp]
\center
\includegraphics[bb=0 0 460 552, width=0.47 \textwidth]{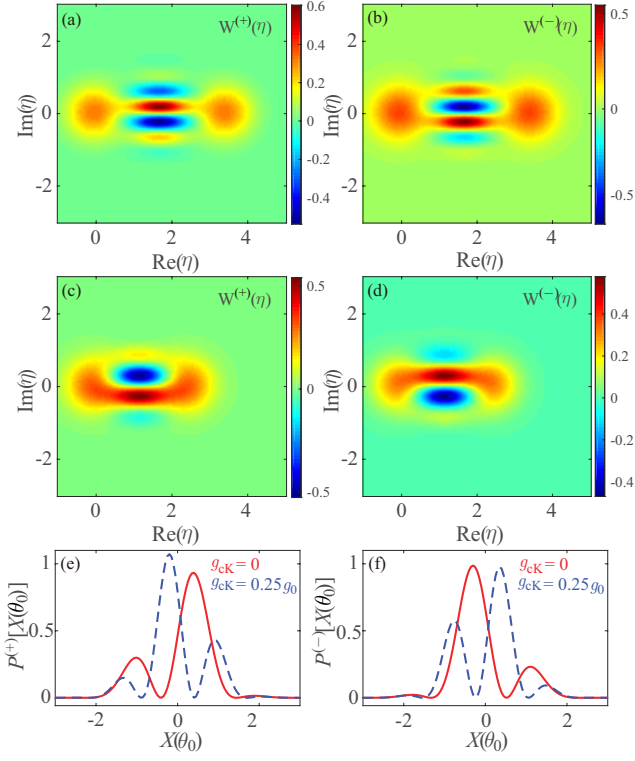}
\caption{(Color online) The Wigner functions $W^{(\pm)}(\eta)$ for the mechanical oscillator states $\vert \Phi^{(\pm)}(t_s)\rangle _{b}$: (a),(b) $g_{\text{cK}}/g_{0}=0.25$ and (c),(d) $g_{\text{cK}}/g_{0}=0$. (e),(f) The probability distributions $P^{(\pm)}[{X}({\theta_0})]$ for the states $\vert\Phi^{(\pm)}(t_s)\rangle_{b}$ as a function of $X({\theta_0})$ at different value of $g_{\text{cK}}/g_0$. The red solid curves correspond to the typical optomechanical system without the cross-Kerr effect ($g_{\text{cK}}/g_0=0$) and The blue dashed curves correspond to the generalized optomechanical system in the presence of the cross-Kerr effect ($g_{\text{cK}}/g_0=0.25$). Other parameters are $\omega_{c}/\omega_{M}=1000$ and $g_{0}/\omega_{M}=1.2$.}
\label{Fig7}
\end{figure}

To explore the quantum coherence and interference effects in the generated mechanical cat states. In Figs.~\ref{Fig7}(a) and~\ref{Fig7}(b), we plot the Wigner functions $W^{(\pm)}(\eta)$ for the mechanical cat states $\vert\Phi^{(\pm)}(t_s)\rangle_{b}$ with $t_s=\pi/(\omega_{M}-g_{\text{cK}})$ being the detection time. Here we can see that the positions of the two main peaks in the Wigner functions are located at the origin and the point corresponding to $\beta(t_s)$ in the phase space, which represent the two coherent states $\vert 0\rangle_{b}$ and $\vert\beta(t_s)\rangle_{b}$. Moreover, we see clear interference pattern (in the region between the two peaks) in the Wigner functions. More importantly, the two main peaks in the Wigner functions of the states $\vert\Phi^{(\pm)}(t_s)\rangle_{b}$ can be distinguished in the phase space, which means that the two superposition components $\vert 0\rangle_{b}$ and $\vert\beta(t_s)\rangle_{b}$ are distinct from each other in the sense of $|_{b}\!\langle0\vert\beta(t_s)\rangle_{b}|\approx0$. We point out that the distinguishability between the two superposition states is enhanced under the assistant of the cross-Kerr interaction. This point can be seen from the expression of the coherence amplitude $|\beta(t_s)|=2g_0/|\omega_{M}-g_{\text{cK}}|$, which increases with the increase of $g_{\text{cK}}$ in the range $g_{\text{cK}}<\omega_{M}$. In addition, in Figs.~\ref{Fig7}(c) and~\ref{Fig7}(d) we show the Wigner functions of the two states in the absence of the cross-Kerr interaction. By comparing the Wigner functions in the two cases: $g_{\text{cK}}/g_{0}=0.25$ and $0$, we can see that the distance between the two peaks is enhanced and that the interference fringes become more clear in the presence of the cross-Kerr interaction. This implies that the cross-Kerr interaction is helpful to the generation of macroscopic mechanical cat state. This enhancement can also be seen from the probability distributions $P^{(\pm)}[{X}({\theta_0})]$ for the states $\vert\Phi^{(\pm)}(t_s)\rangle_{b}$, as shown in Figs.~\ref{Fig7}(e) and~\ref{Fig7}(f). Here, the angle of rotation $\theta_0=\text{arg}[\beta(t_s)]-\pi/2$ is chosen such that the quadrature direction is perpendicular to the link line between the two main peaks. This is because the interference is maximum in this direction due to the probability distributions overlap exactly when the two coherent states are projected onto this quadrature. It can be seen that a stronger oscillation exists in the probability distributions corresponding to the generated cat states in the presence of the cross-Kerr interaction.

\subsection{Effect of the dissipations on the cat state generation}

In this section, we study how the dissipation of the system affects the generation of the cat states. Concretely, we calculate the Wigner function and the probability distribution of $\hat{X}({\theta})$ for the cat states in the presence of dissipation. In this case, the evolution of the system is governed by the quantum master equation~(\ref{SME}) under the replacement of $\hat{H}_{\text{sys}}^{(I)}\rightarrow\hat{H}_{\text{gom}}$. To solve this master equation~(\ref{SME}), we expand the state of the system in the Fock space and write the density matrix as
\begin{eqnarray}
\hat{\rho}(t)=\sum\limits_{m,j,n,k=0}^{\infty}\hat{\rho}_{m,j,n,k}(t)\vert m\rangle_{a}\vert j\rangle_{b}\;_{a}\langle n\vert_{b}\langle k\vert.
\end{eqnarray}
For the initial state $\vert\psi(0)\rangle=(\vert 0\rangle_{a}+\vert 1\rangle_{a})\vert 0\rangle_{b}/\sqrt{2}$, the nonzero density matrix elements are $\hat{\rho}_{0,0,0,0}(0)=\hat{\rho}_{0,0,1,0}(0)=\hat{\rho}_{1,0,0,0}(0)=\hat{\rho}_{1,0,1,0}(0)=1/2$.
By numerically solving the master equation~(\ref{SME}) under the initial condition, the time evolution of the density matrix $\hat{\rho}(t)$ can be obtained.

Accordingly, the density matrices of the mechanical resonator corresponding to the detected cavity states $\vert\pm\rangle_{a}$ can be obtained as
\begin{eqnarray}
\hat{\rho}_{b}^{(\pm)}(t)&=&\frac{_{a}\langle\pm \vert \hat{\rho}(t)\vert\pm\rangle_{a}}{\text{Tr}\left[_{a}\langle\pm\vert\hat{\rho}(t)\vert\pm\rangle_{a}\right]} \notag\\
&=&\frac{1}{2P_{\pm}(t)}\sum\limits_{j,k=0}^{\infty }\Theta_{j,k}^{(\pm)}(t)\vert j\rangle_{b}\;_{b}\langle k\vert , \label{densitymatrices}
\end{eqnarray}
where we introduce the variables
\begin{eqnarray}
\Theta_{j,k}^{(\pm)}(t)=\hat{\rho}_{0,j,0,k}(t)+\hat{\rho}_{1,j,1,k}(t)\pm [\hat{\rho}_{0,j,1,k}(t)+\hat{\rho}_{1,j,0,k}(t)],\nonumber\\
\end{eqnarray}
and the measurement probabilities
\begin{equation}
P_{\pm}(t)=\frac{1}{2}\sum\limits_{j=0}^{\infty }\Theta_{j,j}^{\pm}(t).
\end{equation}

The fidelities between the generate states $\hat{\rho}_{b}^{(\pm)}(t)$ and the target states $\vert \Phi ^{(\pm) }(t) \rangle _{b}$ can also be calculated as
\begin{eqnarray}
F_{\pm}(t)&=&\;_{b}\langle\Phi^{(\pm)}(t)\vert\hat{\rho}_{b}^{(\pm)}(t)\vert\Phi^{(\pm)}(t)\rangle_{b}\nonumber\\
&=&\frac{\mathcal{N}_{\pm}^{2}}{2P_{\pm}(t)}\sum\limits_{j,k=0}^{\infty}\Theta_{j,k}^{(\pm)}(t)\left[\delta_{0,j}\pm e^{-i\vartheta (t)}e^{-\frac{\vert
\beta (t) \vert ^{2}}{2}}\frac{[ \beta ^{\ast }(t)]^{j}}{\sqrt{j!}}\right] \notag \\
&&\times\left[\delta _{k,0}\pm e^{i\vartheta(t)}e^{-\frac{\vert\beta(t)\vert^{2}}{2}}\frac{[\beta(t)]^{k}}{\sqrt{k!}}\right].
\end{eqnarray}
Below, we will analyze the dependence of these fidelities and probabilities on the dissipation parameters: the cavity field decay rate $\kappa$, the mechanical decay rate $\gamma_{M}$, and the average thermal occupation number $\bar{n}_{M}$.
\begin{figure}[tbp]
\center
\includegraphics[bb=0 0 448 439, width=0.47 \textwidth]{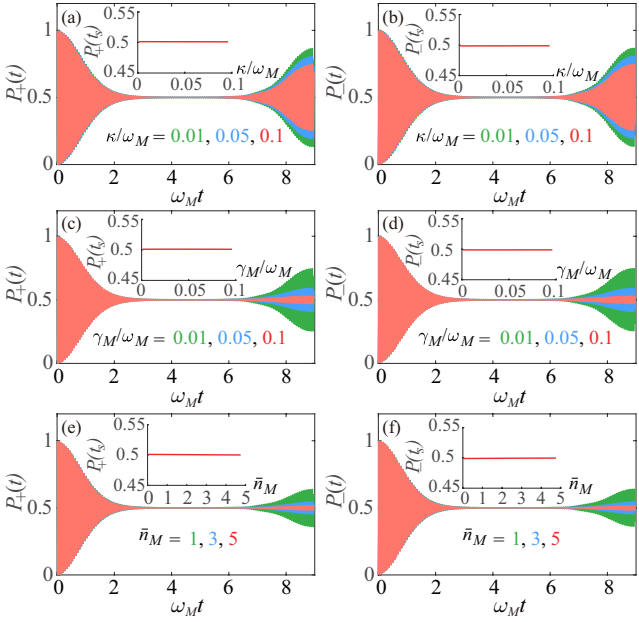}
\caption{(Color online) The probabilities $P_{\pm}(t)$ as a function of time $t$ in various cases. (a),(b) $\gamma_M/\omega_{M}=0.01$, $\bar{n}_{M}=0$, and $\kappa/\omega_{M}= 0.01$, $0.05$, and $0.1$; (c),(d) $\kappa/\omega_{M}=0.1$, $\bar{n}_{M}=0$, and $\gamma_{M}/\omega_{M}=0.01$, $0.05$, and $0.1$; (e),(f) $\kappa/\omega_{M} = 0.1$, $\gamma_{M}/\omega_{M}= 0.01$, and $\bar{n}_{M}= 1$, $3$, and $5$. Other parameters are $\omega_{c}/\omega_{M}=100$, $g_{0}/\omega_{M}=1.2$, and $g_{\text{cK}}/g_{0}=0.25$. The insets are the probabilities $P_{\pm}(t_s)$ at time $t_s=\pi/{(\omega_{M}-g_{ck})}$ vs the dissipation parameters $\kappa/\omega_{M}$, $\gamma_{M}/\omega_{M}$, and $\bar{n}_{M}$.}
\label{Fig8}
\end{figure}

In Fig.~\ref{Fig8}, we display the time dependence of the probabilities $P_{\pm}(t)$ at different values of the cavity-field decay rate $\kappa/\omega_{M}$, the mechanical dissipation rate $\gamma_{M}/\omega_{M}$, and the average thermal phonon number $\bar{n}_{M}$. Figure~\ref{Fig8} shows that the probabilities $P_{\pm}(t)$ oscillate rapidly, which is mainly caused by the free evolution of the cavity field, as shown by the phase factor $\vartheta(t)$ in Eq.~(\ref{varthetabeatt}). With the evolution of the system, the amplitude of the oscillation envelop decreases gradually. In the intermediate duration of $\omega_{M}t\approx3-7$ (around the cavity detection time $t_{s}$), the oscillation amplitude almost disappears and the probabilities $P_{+}(t)\approx P_{-}(t)\approx 1/2$. The amplitude of the oscillation envelop will revive around $t=2\pi/(\omega_{M}-g_{\text{cK}})$ [with the same period as $\beta(t)$]. Comparing to the closed-system case, we see that the amplitude of the oscillation envelop for the probabilities at the revival duration decreases in the presence of dissipations. The insets in Fig.~\ref{Fig8} show the probabilities $P_{\pm}(t_s)$ at $t_s=\pi/{(\omega_{M}-g_{ck})}$ as a function of $\kappa/\omega_{M}$, $\gamma_{M}/\omega_{M}$, and $\bar{n}_{M}$. The results indicate that the probabilities $P_{\pm}(t_s)$ are almost independent of these dissipation parameters.
\begin{figure}[tbp]
\center
\includegraphics[bb=0 0 454 438, width=0.47 \textwidth]{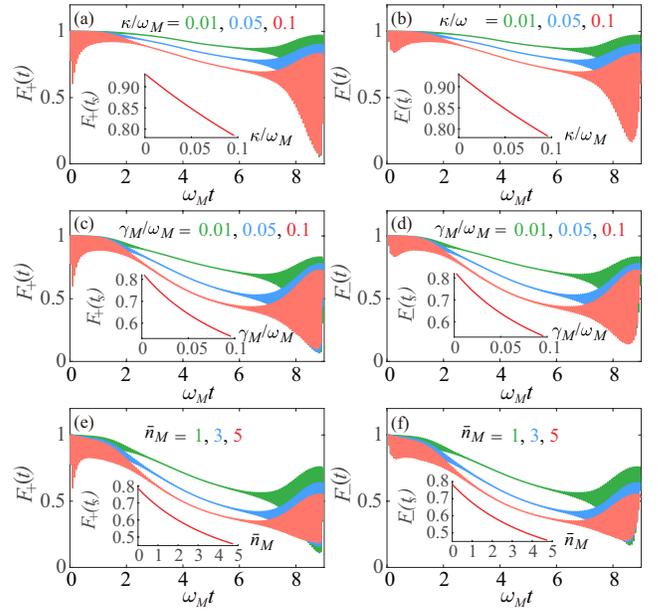}
\caption{(Color online) The fidelities $F_{\pm}(t)$ vs time $t$ in various cases. (a),(b) $\gamma_M/\omega_{M} = 0.01$, $\bar{n}_{M}= 0$, and $\kappa/\omega_{M} = 0.01$, $0.05$, and $0.1$; (c),(d) $\kappa/\omega_{M} = 0.1$, $\bar{n}_{M}= 0$, and $\gamma_M/\omega_{M} = 0.01$, $0.05$, and $0.1$; (e),(f) $\kappa/\omega_{M} = 0.1$, $\gamma_M/\omega_{M} = 0.01$, and $\bar{n}_{M}= 1$, $3$, and $5$. Other parameters are $\omega_{c}/\omega_{M}=100$, $g_{0}/\omega_{M}=1.2$, and $g_{\text{cK}}/g_{0}=0.25$. The insets are the fidelities $F_{\pm}(t_s)$ at time $t_s=\pi/{(\omega_{M}-g_{\text{cK}})}$ vs $\kappa/\omega_{M}$, $\gamma_{M}/\omega_{M}$, and $\bar{n}_{M}$.}
\label{Fig9}
\end{figure}

In Fig.~\ref{Fig9}, we plot the fidelities $F_{\pm}(t)$ as a function of time $t$ at various values of $\kappa/\omega_{M}$, $\gamma_{M}/\omega_{M}$, and $\bar{n}_{M}$. Here we can see that the fidelities exhibit some oscillations at the initial period. With the evolution of the system, the oscillation disappears gradually, and then the oscillation will revive around the time $2\pi/(\omega_{M}-g_{\text{cK}})$. In the intermediate duration of $\omega_{M}t\approx3-7$, the fidelities $F_{+}(t)$ and $F_{-}(t)$ have approximately equal values. The fidelities $F_{\pm}(t)$ have smaller values for larger values of the decay rates $\kappa/\omega_{M}$, $\gamma_{M}/\omega_{M}$, and the thermal occupation number $\bar{n}_{M}$. This phenomenon can also be seen from the insets of Fig.~\ref{Fig9}. Moreover, the plots show that the time dependence of $F_{+}(t)$ is similar to that of $F_{-}(t)$, and that the dependence of $F_{\pm}(t_s)$ on the parameters $\kappa$ ($\gamma_{M}$) and $\bar{n}_{M}$ are almost the same.

The influence of the dissipations on the Wigner functions of the generated state can also be evaluated based on the reduced density matrices $\hat{\rho}_{b}^{(\pm)}(t)$ of the mechanical oscillator. In the open-system case, the Wigner functions of the density matrices $\hat{\rho}_{b}^{(\pm)}(t)$ can be obtained as
\begin{eqnarray}
W^{(\pm)}(\eta)&=&\frac{1}{\pi P_{\pm}(t)}\sum\limits_{l,j,k=0}^{\infty }(-1) ^{l}\Theta_{j,k}^{(\pm)}(t) \notag \\
&&\times _{b}\langle l\vert \hat{D}^{\dagger }(\eta) \vert j\rangle _{b}\;_{b}\langle k\vert \hat{D}(\eta)\vert l\rangle_{b},
\end{eqnarray}
where the matrix elements of the displacement operator in the Fock space can be calculated with Eq.~(\ref{displaopr}).
\begin{figure}[tbp]
\center
\includegraphics[bb=0 0 467 450, width=0.47 \textwidth]{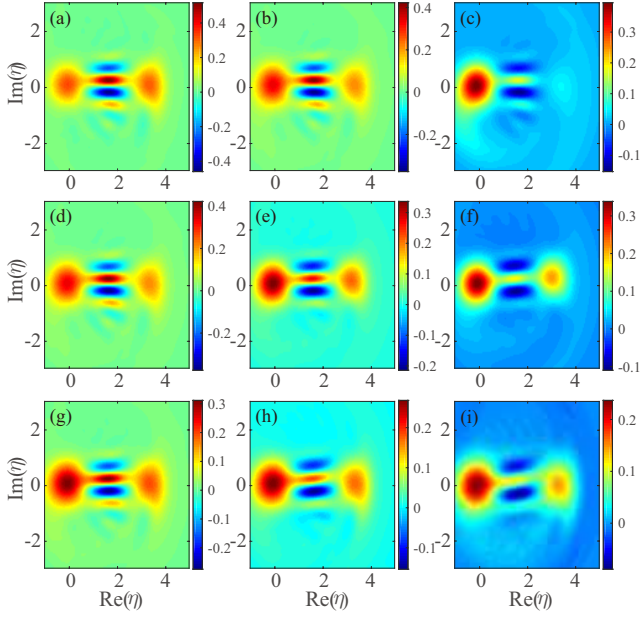}
\caption{(Color online) The Wigner function $W^{(+)}(\eta)$ for the density matrix $\hat{\rho}_{b}^{(+)}(t_s)$ in various cases. (a)-(c) $\gamma_M/\omega_{M} = 0.01$, $\bar{n}_{M}= 0$, and $\kappa/\omega_{M} = 0.01$, $0.1$, and $0.5$; (d)-(f) $\kappa/\omega_{M} = 0.1$, $\bar{n}_{M}= 0$, and $\gamma_M/\omega_{M} = 0.01$, $0.05$, and $0.1$; (g)-(i) $\kappa/\omega_{M} = 0.1$, $\gamma_M/\omega_{M} = 0.01$, and $\bar{n}_{M}= 1$, $3$, and $5$. Other parameters are $\omega_{c}/\omega_{M}=1000$, $g_{0}/\omega_{M}=1.2$, and $g_{\text{cK}}/g_{0}=0.25$.}
\label{Fig10}
\end{figure}

To illustrate how the decay rates and the thermal excitation number of the system affect the Wigner function of the generated mechanical cat states. In Fig.~\ref{Fig10}, the Wigner function $W^{(+)}(\eta)$ for the density matrix $\hat{\rho}_{b}^{(+)}(t_s)$ is plotted when the decay rates of the system and the thermal excitation number take various values. Here we only show the Wigner function $W^{(+) }(\eta)$ for concision because $W^{(-)}(\eta)$ has a similar parameter dependence. We see that, with the increase of the decays rates and the thermal excitation number, the interference patten (in the region between the two peaks) in the Wigner functions attenuates gradually. This means that the decay rates $\kappa$ ($\gamma_{M}$) and the thermal excitation number $\bar{n}_{M}$ of the system fade the macroscopic quantum coherence in the cat states. In addition, with the increase of $\kappa$ ($\gamma_{M}$) and $\bar{n}_{M}$, the peak describing the coherent state $\vert\beta(t)\rangle_{b}$ in the Wigner function reduces gradually.

We also study the effect of the dissipation on the probability distributions of the rotated quadrature operator $\hat{X}({\theta})$. In this case, the probability distributions of $\hat{X}({\theta})$ can be obtained as
\begin{eqnarray}
P^{(\pm)}[X({\theta})]
&=& \frac{e^{-X^2(\theta)}}{2P_{\pm}(t)}\sum\limits_{j,k=0}^{\infty }\frac{\Theta_{j,k}^{(\pm)}(t)}{\sqrt{\pi2^{j+k}j!k!}}\notag \\
&& \times H_{j}[X({\theta})]H_{k}[X({\theta})]e^{i\theta (k-j)}.
\end{eqnarray}

\begin{figure}[tbp]
\center
\includegraphics[bb=0 0 451 448, width=0.47 \textwidth]{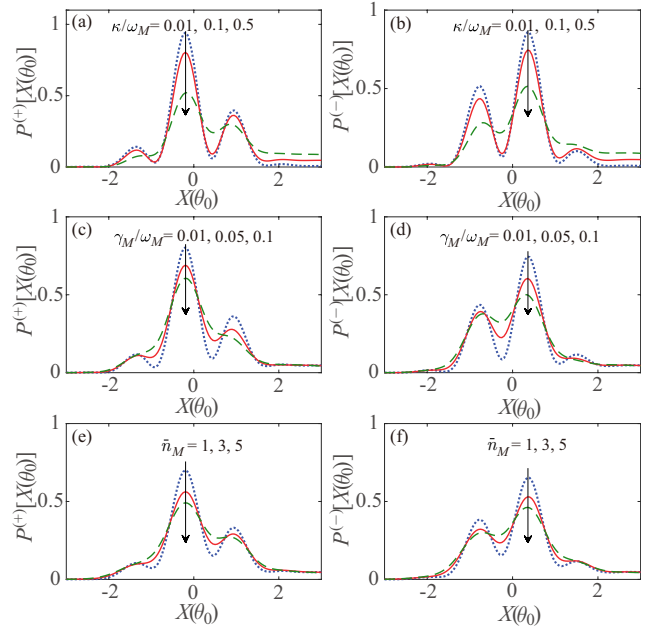}
\caption{(Color online) The probability distribution $P^{(\pm)}[{X}({\theta_0})]$ for the density matrices $\hat{\rho}_{b}^{(\pm)}(t_s)$ as a function of $X({\theta_0})$ in various cases: (a),(b) $\gamma_M/\omega_{M} = 0.01$, $\bar{n}_{M}= 0$, and $\kappa/\omega_{M}= 0.01$, $0.1$, and $0.5$; (c),(d) $\kappa/\omega_{M} = 0.1$, $\bar{n}_{M}= 0$, and $\gamma_M/\omega_{M} = 0.01$, $0.05$, and $0.1$; (e),(f) $\kappa/\omega_{M} = 0.1$, $\gamma_M/\omega_{M} = 0.01$, and $\bar{n}_{M} = 1$, $3$, and $5$. Other parameters are $\omega_{c}/\omega_{M}=1000$, $g_{0}/\omega_{M}=1.2$, and $g_{\text{cK}}/g_{0}=0.25$.}
\label{Fig11}
\end{figure}

In Fig.~\ref{Fig11}, we plot the probability distributions $P^{(\pm)}[{X}({\theta_0})]$ for the density matrices $\hat{\rho}_{b}^{(\pm)}(t_s)$ as a function of $X({\theta_0})$ when the decay rates of the system and the thermal excitation number take various values. It can be seen that, with the increase of the decay rates and the thermal excitation number, the oscillation amplitude of the probability distributions decreases gradually, which means that the decay rates and the thermal excitation number of the system hurt the macroscopic quantum coherence.

\section{Discussions and conclusion \label{conclusion}}

We now present some discussions on the experimental parameters for implementation of this model. In principle, the studies in this work are general and it can be implemented with various optomechanical systems which can be described by this generalized optomechanical model. Below, we focus our experimental analyses on a superconducting circuit because this generalized optomechanical model has been proposed to enhance the single-photon optomechanical coupling in this setup~\cite{Heikkila2014PRL}. In particular, this coupling enhancement scheme has recently been realized in a superconducting circuit~\cite{Pirkkalainen2015NC}.
Nevertheless, we should point out that some used parameters are accessible with current experiments, but there still exists some challenge for current experimental technology. For observation of the photon blockade effect, the system is expected to work in the single-photon strong-coupling regime $g_{0}>\kappa$. For generation of the cat states, the state generation time $t_{s}=\pi/(\omega_{M}-g_{\text{cK}})$ is required to be shorter than the lifetime $1/\kappa$ of the cavity photon, which leads to the resolved-sideband condition $\omega_{M}\gg\kappa$. In our simulations, we used these parameters: $g_{0}/\omega_{M}\approx0.5$ - $1.2$, $g_{\text{cK}}/g_{0}\approx0.25$, $\kappa/\omega_{M}\approx0.01$- $0.5$, and $\gamma_{M}/\omega_{M}\approx0.001$ - $0.1$. These used parameters have been evaluated to be accessible with the near-future technology~\cite{Heikkila2014PRL,Pirkkalainen2015NC}. For example, when the mechanical resonance frequency is taken as $\omega_{M}\sim2\pi\times10$ MHz, the optomechanical coupling strength is $g_{0}\sim2\pi\times5$ - $12$ MHz. Note that a coupling strength of the order of $g_{0}\sim2\pi\times100$ MHz has been evaluated to be in principle possible with an optimized device~\cite{Pirkkalainen2015NC}. The cross-Kerr interaction strength $g_{\text{cK}}/g_{0}\approx0.25$ has also been estimated in this system~\cite{Heikkila2014PRL}. In addition, the cavity decay rate (of the order of $\sim1$ MHz) and the mechanical decay rate ($\sim100$ kHz) are accessible with the current experimental conditions~\cite{Aspelmeyer2014RMP}.

In conclusion, we have studied the few-photon optomechanical effects in a generalized optomechanical system in which there exist both the optomechanical coupling and the cross-Kerr interaction between the optical and phononic modes. In particular, we focused on the photon blockade effect and the generation of the mechanical cat states in both the close- and open-system cases. We found that the cross-Kerr interaction can strengthen or attenuate the photon blockade of the cavity by calculating the equal-time second-order correlation function. We also found that the cross-Kerr interaction can enhance the quantum interference and coherence effect in the generated mechanical cat states by calculating the Wigner function and the probability distribution of the rotated quadrature operator.

\begin{acknowledgments}
J.-F.H. is supported in part by National Natural Science Foundation of China under Grant No.~11505055. J.-Q.L. is supported in part by National Natural Science Foundation of China under Grants No.~11822501 and No.~11774087, and Hunan Provincial Natural Science Foundation of China under Grant No.~2017JJ1021.
\end{acknowledgments}

\appendix*
\begin{widetext}

\section{Derivation of the unitary evolution operator $\hat{U}(t)$ \label{appendix}}

In this Appendix, we present a detailed derivation of the unitary evolution operator $\hat{U}(t)$ given in Eq.~(\ref{uniopr}). For the Hamiltonian $\hat{H}_{\text{gom}}$, its unitary evolution operator $\hat{U}(t)$ can be expressed as
\begin{eqnarray}
\hat{U}(t)&=&e^{-i\omega_{c}t\hat{a}^{\dagger}\hat{a}-i\omega_{M}t\hat{b}^{\dagger}\hat{b}+ig_{0}t\hat{a}^{\dagger}\hat{a}(\hat{b}^{\dagger}+\hat{b})+ig_{\text{cK}}t\hat{a}^{\dagger}\hat{a}\hat{b}^{\dagger}\hat{b}}.
\end{eqnarray}
To decompose this unitary operator, we introduce a unitary transformation defined by $\hat{D}(\hat{\xi})=\exp[\hat{\xi}(\hat{b}^{\dagger}-\hat{b})]$, where $\hat{\xi}$ is defined in Eq.~(\ref{definbeta}). Using the Baker-Cambell-Hausdorf expansion~\cite{Louisell}, the transformed operator can be obtained as
\begin{eqnarray}
\hat{D}^{\dagger}(\hat{\xi})\hat{U}(t) \hat{D}(\hat{\xi})&=&e^{-i\omega _{c}t\hat{a}^{\dagger }\hat{a}}e^{i( g_{\text{cK}}\hat{a}^{\dagger }\hat{a}-\omega
_{M})t\hat{b}^{\dagger }\hat{b}}e^{i\left[ \frac{2g_{0}^{2}}{\omega
_{M}-g_{\text{cK}}\hat{a}^{\dagger }\hat{a}}-\frac{\omega _{M}g_{0}^{2}}{\left( \omega
_{M}-g_{\text{cK}}\hat{a}^{\dagger }\hat{a}\right) ^{2}}\right] t\hat{a}^{\dagger }\hat{a}\hat{a}^{\dagger }\hat{a}}e^{
\frac{ig_{\text{cK}}g_{0}^{2}t}{\left( \omega _{M}-g_{\text{cK}}\hat{a}^{\dagger }\hat{a}\right) ^{2}}
\hat{a}^{\dagger }\hat{a}\hat{a}^{\dagger }\hat{a}\hat{a}^{\dagger }\hat{a}}.
\end{eqnarray}
Then we can obtain the following expression for the unitary evolution operator
\begin{eqnarray}
\hat{U}(t) &=&\hat{D}(\hat{\xi})e^{-i\omega _{c}t\hat{a}^{\dagger }\hat{a}}e^{i(g_{\text{cK}}\hat{a}^{\dagger }\hat{a}-\omega _{M})t\hat{b}^{\dagger }\hat{b}}e^{i\left[ \frac{%
2g_{0}^{2}}{\omega _{M}-g_{\text{cK}}\hat{a}^{\dagger }\hat{a}}-\frac{\omega _{M}g_{0}^{2}}{(\omega _{M}-g_{\text{cK}}\hat{a}^{\dagger }\hat{a}) ^{2}}\right] t\hat{a}^{\dagger
}\hat{a}\hat{a}^{\dagger }\hat{a}}e^{\frac{ig_{\text{cK}}g_{0}^{2}t}{( \omega
_{M}-g_{\text{cK}}\hat{a}^{\dagger }\hat{a})^{2}}\hat{a}^{\dagger }\hat{a}\hat{a}^{\dagger }\hat{a}\hat{a}^{\dagger
}\hat{a}}\hat{D}^{\dagger}(\hat{\xi}) \notag \\
&=&e^{-i\omega _{c}t\hat{a}^{\dagger }\hat{a}}e^{i\left[ \frac{2g_{0}^{2}}{\omega
_{M}-g_{\text{cK}}\hat{a}^{\dagger }\hat{a}}-\frac{\omega _{M}g_{0}^{2}}{( \omega_{M}-g_{\text{cK}}\hat{a}^{\dagger }\hat{a}) ^{2}}\right] t\hat{a}^{\dagger }\hat{a}\hat{a}^{\dagger }\hat{a}}e^{\frac{ig_{\text{cK}}g_{0}^{2}t}{(\omega _{M}-g_{\text{cK}}\hat{a}^{\dagger }\hat{a})^{2}}
\hat{a}^{\dagger }\hat{a}\hat{a}^{\dagger }\hat{a}\hat{a}^{\dagger }\hat{a}} \notag \\
&&\times e^{\frac{g_{0}}{\omega _{M}-g_{\text{cK}}\hat{a}^{\dagger }\hat{a}}\hat{a}^{\dagger }\hat{a}(\hat{b}^{\dagger }-\hat{b}) }e^{i( g_{\text{cK}}\hat{a}^{\dagger }\hat{a}-\omega _{M})
t\hat{b}^{\dagger }\hat{b}}e^{-\frac{g_{0}}{\omega_{M}-g_{\text{cK}}\hat{a}^{\dagger }\hat{a}}\hat{a}^{\dagger}\hat{a}( \hat{b}^{\dagger }-\hat{b})}.\label{eqA3}
\end{eqnarray}
Based on the relation
\begin{eqnarray}
e^{i(g_{\text{cK}}\hat{a}^{\dagger }\hat{a}-\omega _{M}) t\hat{b}^{\dagger }\hat{b}}e^{-\frac{
g_{0}}{\omega _{M}-g_{\text{cK}}\hat{a}^{\dagger }\hat{a}}\hat{a}^{\dagger }\hat{a}( \hat{b}^{\dagger}-\hat{b}) }e^{-i(g_{\text{cK}}\hat{a}^{\dagger }\hat{a}-\omega _{M}) t\hat{b}^{\dagger
}\hat{b}} &=&e^{-\frac{g_{0}}{\omega _{M}-g_{\text{cK}}\hat{a}^{\dagger }\hat{a}}\hat{a}^{\dagger }\hat{a}\left[
\hat{b}^{\dagger }e^{i(g_{\text{cK}}\hat{a}^{\dagger }\hat{a}-\omega _{M})
t}-\hat{b}e^{-i( g_{\text{cK}}\hat{a}^{\dagger }\hat{a}-\omega _{M})t}\right]},
\end{eqnarray}
we obtain
\begin{eqnarray}
e^{i(g_{\text{cK}}\hat{a}^{\dagger }\hat{a}-\omega _{M}) t\hat{b}^{\dagger }\hat{b}}e^{-\frac{g_{0}}{\omega _{M}-g_{\text{cK}}\hat{a}^{\dagger }\hat{a}}\hat{a}^{\dagger }\hat{a}( \hat{b}^{\dagger
}-\hat{b}) } &=&e^{-\frac{g_{0}}{\omega _{M}-g_{\text{cK}}\hat{a}^{\dagger }\hat{a}}\hat{a}^{\dagger
}\hat{a}\left[ \hat{b}^{\dagger }e^{i( g_{\text{cK}}\hat{a}^{\dagger }\hat{a}-\omega _{M})
t}-\hat{b}e^{-i( g_{\text{cK}}\hat{a}^{\dagger }\hat{a}-\omega _{M}) t}\right] }e^{i(g_{\text{cK}}\hat{a}^{\dagger }\hat{a}-\omega _{M})t\hat{b}^{\dagger }\hat{b}}.\label{eqA5}
\end{eqnarray}
By inserting Eq.~(\ref{eqA5}) into Eq.~(\ref{eqA3}) and using the following relation
\begin{eqnarray}
&&e^{\frac{g_{0}}{\omega_{M}-g_{\text{cK}}\hat{a}^{\dagger}\hat{a}}\hat{a}^{\dagger}\hat{a}(\hat{b}^{\dagger}-\hat{b})}e^{-\frac{g_{0}}{\omega_{M}-g_{\text{cK}}\hat{a}^{\dagger}\hat{a}}
\hat{a}^{\dagger}\hat{a}[\hat{b}^{\dagger}e^{i(g_{\text{cK}}\hat{a}^{\dagger}\hat{a}-\omega_{M})t}-\hat{b}e^{-i(g_{\text{cK}}\hat{a}^{\dagger}\hat{a}-\omega_{M})t}]} \notag \\
&=&e^{-i\frac{g_{0}^{2}}{( \omega _{M}-g_{\text{cK}}\hat{a}^{\dagger }\hat{a})^{2}}\sin[(\omega_{M}-g_{\text{cK}}\hat{a}^{\dagger}\hat{a})t]\hat{a}^{\dagger }\hat{a}\hat{a}^{\dagger }\hat{a}}e^{\frac{g_{0}}{\omega _{M}-g_{\text{cK}}\hat{a}^{\dagger}\hat{a}}
\hat{a}^{\dagger }\hat{a}[(1-e^{i(g_{\text{cK}}\hat{a}^{\dagger}\hat{a}-\omega_{M})t}) \hat{b}^{\dagger}-(1-e^{-i(g_{\text{cK}}\hat{a}^{\dagger}\hat{a}-\omega_{M})t})\hat{b}]}.
\end{eqnarray}
we obtain the unitary operator as
\begin{eqnarray}
\hat{U}(t)&=&e^{-i\omega_{c}t\hat{a}^{\dagger}\hat{a}}e^{i\frac{g_{0}^{2}}{(\omega_{M}-g_{\text{cK}}\hat{a}^{\dagger}\hat{a})^{2}}
[\omega_{M}t-\sin(\omega_{M}t-g_{\text{cK}}\hat{a}^{\dagger}\hat{a}t)]\hat{a}^{\dagger }\hat{a}\hat{a}^{\dagger}\hat{a}}e^{\frac{-ig_{\text{cK}}
g_{0}^{2}t}{(\omega_{M}-g_{\text{cK}}\hat{a}^{\dagger}\hat{a})^{2}}\hat{a}^{\dagger}\hat{a}\hat{a}^{\dagger}\hat{a}\hat{a}^{\dagger}\hat{a}}\notag\\
&&\times e^{\frac{g_{0}}{\omega_{M}-g_{\text{cK}}\hat{a}^{\dagger}\hat{a}}\hat{a}^{\dagger}\hat{a}[(1-e^{i(g_{\text{cK}}\hat{a}^{\dagger}\hat{a}-\omega_{M})t})
\hat{b}^{\dagger }-(1-e^{-i(g_{\text{cK}}\hat{a}^{\dagger}\hat{a}-\omega_{M})t})\hat{b}] }e^{i(g_{\text{cK}}\hat{a}^{\dagger }\hat{a}-\omega _{M})t\hat{b}^{\dagger }\hat{b}},
\end{eqnarray}
which can be further expressed as
\begin{equation}
\hat{U}(t)=e^{-i\omega_{c}t\hat{a}^{\dagger}\hat{a}}e^{i\hat{\mu}(t)\hat{a}^{\dagger}\hat{a}\hat{a}^{\dagger}\hat{a}}e^{-i\hat{\nu}(t)\hat{a}^{\dagger}\hat{a}\hat{a}^{\dagger}\hat{a}\hat{a}^{\dagger}\hat{a}}
e^{\hat{a}^{\dagger}\hat{a}[\hat{\lambda}(t)\hat{b}^{\dagger}-\hat{\lambda}^{\ast}(t)\hat{b}]}e^{i(g_{\text{cK}}\hat{a}^{\dagger}\hat{a}-\omega_{M})\hat{b}^{\dagger}\hat{b}t},
\end{equation}
where the variables $ \hat{\mu}(t)$, $\hat{\nu}(t)$, and $\hat{\lambda}(t)$ have been given by Eqs.~(\ref{uniopevariab}).
\end{widetext}


\begin{thebibliography}{99}

\bibitem{Vahala2008Science} T. J. Kippenberg and K. J. Vahala, Cavity Optomechanics: Back-Action at the Mesoscale, Science \textbf{321}, 1172 (2008).
\bibitem{Meystre2012PT} M. Aspelmeyer, P. Meystre, and K. Schwab, Quantum optomechanics, Phys. Today \textbf{65}, 29 (2012).
\bibitem{Aspelmeyer2014RMP} M. Aspelmeyer, T. J. Kippenberg, and F. Marquardt, Cavity optomechanics, Rev. Mod. Phys. \textbf{86}, 1391 (2014).

\bibitem{Law1995PRA} C. K. Law, Interaction between a moving mirror and radiation pressure: A Hamiltonian formulation, Phys. Rev. A \textbf{51}, 2537 (1995).

\bibitem{Dobrindt2008PRL}   J. M. Dobrindt, I. Wilson-Rae, and T. J. Kippenberg, Parametric Normal-Mode Splitting in Cavity Optomechanics, Phys. Rev. Lett. \textbf{101}, 263602 (2008).
\bibitem{Groblacher2009Nature}     S. Gr\"{o}blacher, K. Hammerer, M. R. Vanner, and M. Aspelmeyer, Observation of strong coupling between a micromechanical resonator and an optical cavity field, Nature (London) \textbf{460}, 724 (2009).
\bibitem{Teufel2011NatureA}     J. D. Teufel, D. Li, M. S. Allman, K. Cicak, A. J. Sirois, J. D. Whittaker, and R. W. Simmonds, Circuit cavity electromechanics in the strong-coupling regime, Nature (London), \textbf{471}, 204 (2011).
\bibitem{Verhagen2012Nature}   E. Verhagen, S. Del\'{e}glise, S. Weis, A. Schliesser, and T. J. Kippenberg, Quantum-coherent coupling of a mechanical oscillator to an optical cavity mode, Nature (London) \textbf{482}, 63 (2012).

\bibitem{Wilson-Rae2007PRL}       I. Wilson-Rae, N. Nooshi, W. Zwerger, and T. J. Kippenberg, Theory of Ground State Cooling of a Mechanical Oscillator Using Dynamical Backaction, Phys. Rev. Lett. \textbf{99}, 093901 (2007).
\bibitem{Marquardt2007PRL}         F. Marquardt, J. P. Chen, A. A. Clerk, and S. M. Girvin, Quantum Theory of Cavity-Assisted Sideband Cooling of Mechanical Motion, Phys. Rev. Lett. \textbf{99}, 093902 (2007).
\bibitem{Genes2008PRA}               C. Genes, D. Vitali, P. Tombesi, S. Gigan, and M. Aspelmeyer, Ground-state cooling of a micromechanical oscillator: Comparing cold damping and cavity-assisted cooling schemes, Phys. Rev. A \textbf{77}, 033804 (2008); Phys. Rev. A \textbf{79}, 039903(E) (2009).
\bibitem{Teufel2011Nature}   J. D. Teufel, T. Donner, D. Li, J. W. Harlow, M. S. Allman, K. Cicak, A. J. Sirois, J. D. Whittaker, K. W. Lehnert, and R. W. Simmonds, Sideband cooling of micromechanical motion to the quantum ground state, Nature \textbf{475}, 359 (2011).
\bibitem{Chan2011Nature}     J. Chan, T. P. M. Alegre, A. H. Safavi-Naeini, J. T. Hill, A. Krause, S. Gr\"{o}blacher, M. Aspelmeyer, and O. Painter, Laser cooling of a nanomechanical oscillator into its quantum ground state, Science \textbf{478}, 89 (2011).

\bibitem{Li2008PRB}       Y. Li, Y.-D. Wang, F. Xue, and C. Bruder, Quantum theory of transmission line resonator-assisted cooling of a micromechanical resonator, Phys. Rev. B \textbf{78}, 134301 (2008).
\bibitem{Liu2013PRL}       Y.-C. Liu, Y.-F. Xiao, X. Luan, and C. W. Wong, Dynamic Dissipative Cooling of a Mechanical Resonator in Strong Coupling Optomechanics, Phys. Rev. Lett. \textbf{110}, 153606 (2013).
\bibitem{Lai2018PRA}           D.-G. Lai, F. Zou, B. P. Hou, Y.-F. Xiao, and J.-Q. Liao, Simultaneous cooling of coupled mechanical resonators in cavity optomechanics, Phys. Rev. A \textbf{98}, 023860 (2018).

\bibitem{Vitali2007PRL}       D. Vitali, S. Gigan, A. Ferreira, H. R. B\"{o}hm, P. Tombesi, A. Guerreiro, V. Vedral, A. Zeilinger, and M. Aspelmeyer, Optomechanical Entanglement between a Movable Mirror and a Cavity Field, Phys. Rev. Lett. \textbf{98}, 030405 (2007).
\bibitem{Hartmann2008PRL}    M. J. Hartmann and M. B. Plenio, Steady State Entanglement in the Mechanical Vibrations of Two Dielectric Membranes, Phys. Rev. Lett. \textbf{101}, 200503 (2008).
\bibitem{Tian2013PRL}       L. Tian, Robust Photon Entanglement via Quantum Interference in Optomechanical Interfaces, Phys. Rev. Lett. \textbf{110}, 233602 (2013).
\bibitem{Wang2013PRL}     Y.-D. Wang and A. A. Clerk, Reservoir-Engineered Entanglement in Optomechanical Systems, Phys. Rev. Lett. \textbf{110}, 253601 (2013).
\bibitem{Palomaki2013Science}       T. A. Palomaki, J. D. Teufel, R. W. Simmonds, and K. W. Lehnert, Entangling Mechanical Motion with Microwave Fields, Science \textbf{342}, 710 (2013).

\bibitem{Agarwal2010PRA}       G. S. Agarwal and S. Huang, Electromagnetically induced transparency in mechanical effects of light, Phys. Rev. A \textbf{81}, 041803(R) (2010).
\bibitem{Weis2010Science}       S. Weis, R. Rivi\`{e}re, S. Del\'{e}glise, E. Gavartin, O. Arcizet, A. Schliesser, and T. J. Kippenberg, Optomechanically Induced Transparency, Science \textbf{330}, 1520 (2010).
\bibitem{Safavinaeini2011Nature}       A. H. Safavinaeini, T. P. M. Alegre, J. Chan, M. Eichenfield, M. Winger, Q. Lin, J. T. Hill, D. E. Chang, and O. Painter, Electromagnetically induced transparency and slow light with optomechanics, Nature (London) \textbf{472}, 69 (2011).

\bibitem{Brooks2012Nature}     D. W. C. Brooks, T. Botter, S. Schreppler, T. P. Purdy, N. Brahms, and D. M. Stamper-Kurn, Non-classical light generated by quantum-noise-driven cavity optomechanics, Nature (London) \textbf{488}, 476 (2012).
\bibitem{Naeini2013Nature}    A. H. Safavi-Naeini, S. Gr\"{o}blacher, J. T. Hill, J. Chan, M. Aspelmeyer, and O. Painter, Squeezed light from a silicon micromechanical resonator, Nature (London) \textbf{500}, 185 (2013).
\bibitem{Purdy2013PRX}   T. P. Purdy, P.-L. Yu, R. W. Peterson, N. S. Kampel, and C. A. Regal, Strong Optomechanical Squeezing of Light, Phys. Rev. X \textbf{3}, 031012 (2013).

\bibitem{Rabl2011PRL} P. Rabl, Photon Blockade Effect in Optomechanical Systems, Phys. Rev. Lett. \textbf{107}, 063601 (2011).
\bibitem{Nunnenkamp2011PRL} A. Nunnenkamp, K. B{\o}rkje, and S. M. Girvin, Single-Photon Optomechanics, Phys. Rev. Lett. \textbf{107}, 063602 (2011).
\bibitem{Liao2012PRA} J.-Q. Liao, H. K. Cheung, and C. K. Law, Spectrum of single-photon emission and scattering in cavity optomechanics, Phys. Rev. A \textbf{85}, 025803 (2012).
\bibitem{Hong2013PRA}  T. Hong, H. Yang, H. Miao, and Y. Chen, Open quantum dynamics of single-photon optomechanical devices, Phys. Rev. A \textbf{88}, 023812 (2013).
\bibitem{Liao2013PRA} J.-Q. Liao and C. K. Law, Correlated two-photon scattering in cavity optomechanics, Phys. Rev. A \textbf{87}, 043809 (2013).
\bibitem{Liu2013PRA} X.-W. Xu, Y.-J. Li, and Y.-x. Liu, Photon-induced tunneling in optomechanical systems, Phys. Rev. A \textbf{87}, 025803 (2013).

\bibitem{Imamoglu1997PRL} A. Imamo\v{g}lu, H. Schmidt, G. Woods, and M. Deutsch, Strongly Interacting Photons in a Nonlinear Cavity, Phys. Rev. Lett. \textbf{79}, 1467 (1997).
\bibitem{Birnbaum2005nature} K. M. Birnbaum, A. Boca, R. Miller, A. D. Boozer, T. E. Northup, and H. J. Kimble, Photon blockade in an optical cavity with one trapped atom, Nature (London) \textbf{436}, 87 (2005).
\bibitem{Liew2010PRL} T. C. H. Liew and V. Savona, Single Photons from Coupled Quantum Modes, Phys. Rev. Lett. \textbf{104}, 183601 (2010).
\bibitem{Bamba2011PRA} M. Bamba, A. Imamo\v{g}lu, I. Carusotto, and C. Ciuti, Origin of strong photon antibunching in weakly nonlinear photonic molecules, Phys. Rev. A \textbf{83}, 021802(R) (2011).
\bibitem{Nori2013PRA} J.-Q. Liao and F. Nori, Photon blockade in quadratically coupled optomechanical systems, Phys. Rev. A \textbf{88}, 023853 (2013).
\bibitem{Huang2013PRA} J.-F. Huang, J.-Q. Liao, and C. P. Sun, Photon blockade induced by atoms with Rydberg coupling, Phys. Rev. A \textbf{87}, 023822 (2013).
\bibitem{Huang2018PRL}  R. Huang, A. Miranowicz, J.-Q. Liao, F. Nori, and H. Jing, Nonreciprocal Photon Blockade, Phys. Rev. Lett. \textbf{121}, 153601 (2018).

\bibitem{Marshall2003PRL} W. Marshall, C. Simon, R. Penrose, and D. Bouwmeester, Towards Quantum Superpositions of a Mirror, Phys. Rev. Lett. \textbf{91}, 130401 (2003).
\bibitem{Tian2005PRB} L. Tian, Entanglement from a nanomechanical resonator weakly coupled to a single Cooper-pair box, Phys. Rev. B \textbf{72}, 195411 (2005).
\bibitem{Isart2011PRL} O. Romero-Isart, A. C. Pflanzer, F. Blaser, R. Kaltenbaek, N. Kiesel, M. Aspelmeyer, and J. I. Cirac, Large Quantum Superpositions and Interference of Massive Nanometer-Sized Objects, Phys. Rev. Lett. \textbf{107}, 020405 (2011).
\bibitem{Pepper2012PRL} B. Pepper, R. Ghobadi, E. Jeffrey, C. Simon, and D. Bouwmeester, Optomechanical Superpositions via Nested Interferometry, Phys. Rev. Lett. \textbf{109}, 023601 (2012).
\bibitem{Yin2013PRA} Z.-q. Yin, T. Li, X. Zhang, and L. M. Duan, Large quantum superpositions of a levitated nanodiamond through spin-optomechanical coupling, Phys. Rev. A \textbf{88}, 033614 (2013).
\bibitem{Tan2013PRA} H. Tan, F. Bariani, G. Li, and P. Meystre, Generation of macroscopic quantum superpositions of optomechanical oscillators by dissipation, Phys. Rev. A \textbf{88}, 023817 (2013).
\bibitem{Ge2015PRA} W. Ge and M. S. Zubairy, Macroscopic optomechanical superposition via periodic qubit flipping, Phys. Rev. A \textbf{91}, 013842 (2015).
\bibitem{Liao2016PRA} J.-Q. Liao, J.-F. Huang, and L. Tian, Generation of macroscopic Schr\"{o}dinger-cat states in qubit-oscillator systems, Phys. Rev. A \textbf{93}, 033853 (2016).
\bibitem{Liao2016PRL} J.-Q. Liao and L. Tian, Macroscopic Quantum Superposition in Cavity Optomechanics, Phys. Rev. Lett. \textbf{116}, 163602 (2016).

\bibitem{Liao2015PRA}    J.-Q. Liao, C. K. Law, L.-M. Kuang, and F. Nori, Enhancement of mechanical effects of single photons in modulated two-mode optomechanics, Phys. Rev. A \textbf{92}, 013822 (2015).

\bibitem{Xuereb2012PRL}      A. Xuereb, C. Genes, and A. Dantan, Strong Coupling and Long-Range Collective Interactions in Optomechanical Arrays, Phys. Rev. Lett. \textbf{109}, 223601 (2012).
\bibitem{Rimberg2014NJP}    A. J. Rimberg, M. P. Blencowe, A. D. Armour, and P. D. Nation, A cavity-Cooper pair transistor scheme for investigating quantum optomechanics in the ultra-strong coupling regime, New J. Phys. \textbf{16}, 055008 (2014).
\bibitem{Heikkila2014PRL}   T. T. Heikkil\"{a}, F. Massel, J. Tuorila, R. Khan, and M. A. Sillanp\"{a}\"{a}, Enhancing Optomechanical Coupling via the Josephson Effect, Phys. Rev. Lett. \textbf{112}, 203603 (2014).

\bibitem{Pirkkalainen2015NC}   J.-M. Pirkkalainen, S. U. Cho, F. Massel, J. Tuorila, T. T. Heikkil\"{a}, P. J. Hakonen, and M. A. Sillanp\"{a}\"{a}, Cavity optomechanics mediated by a quantum two-level system, Nat. Commun. \textbf{6}, 6981 (2015).

\bibitem{Liao2014NJP}   J.-Q. Liao, K. Jacobs, F. Nori, and R. W. Simmonds, Modulated electromechanics: large enhancements of nonlinearities, New J. Phys. \textbf{16}, 072001 (2014).
\bibitem{Lue2015PRL}    X.-Y. L\"{u}, Y. Wu, J. R. Johansson, H. Jing, J. Zhang, and F. Nori, Squeezed Optomechanics with Phase-Matched Amplification and Dissipation, Phys. Rev. Lett. \textbf{114}, 093602 (2015).
\bibitem{Lemonde2016NC}   M.-A. Lemonde, N. Didier, and A. A. Clerk, Enhanced nonlinear interactions in quantum optomechanics via mechanical amplification, Nat. Commun. \textbf{7}, 11338 (2016).

\bibitem{Heikkila2015PRA} R. Khan, F. Massel, and T. T. Heikkil\"{a}, Cross-Kerr nonlinearity in optomechanical systems, Phys. Rev. A \textbf{91}, 043822 (2015).
\bibitem{You2016PRA} W. Xiong, D.-Y. Jin, Y. Qiu, C.-H. Lam, and J. Q. You, Cross-Kerr effect on an optomechanical system, Phys. Rev. A \textbf{93}, 023844 (2016).
\bibitem{Sarma2017JOS} S. Chakraborty and A. K. Sarma, Enhancing quantum correlations in an optomechanical system via cross-Kerr nonlinearity, J. Opt. Soc. Am. B \textbf{34}, 1503 (2017).

\bibitem{Oliveira1990PRA} F. A. M. de Oliveira, M. S. Kim, P. L. Knight, and V. Bu\v{z}ek, Properties of displaced number states, Phys. Rev. A \textbf{41}, 2645 (1990).
\bibitem{Buzek} V. Bu\v{z}ek and P. L. Knight, in \emph{Quantum interference, superposition States of Light, and Nonclassical Effects}, Progress in Optics Vol. XXXIV, edited by E. Wolf (Elsevier, Amsterdam, 1995).
\bibitem{Walls} D. F. Walls and G. J. Millburn, \emph{Quantum Optics} (Springer, Berlin, 2008).
\bibitem{Louisell} W. H. Louisell, \emph{Quantum Statistical Properties of Radiation} (John Wiley $\&$ Sons, New York, 1990).
\end{thebibliography}
\end{document}